\documentclass[a4paper,12pt]{article}
\usepackage{jheppub} 
\usepackage{lineno}
\allowdisplaybreaks[3]

\title{\boldmath Accelerating Unruh-DeWitt detectors coupled with a spinor field}

\author[a]{Dawei Wu}
\author[a]{Shan-Chang Tang}
\author[b,c,a,1]{Yu Shi \note{Corresponding author.}}
\affiliation[a]{Department of Physics \& State Key Laboratory of Surface Physics, Fudan University, Shanghai 200438, China}
\affiliation[b]{University of Science and Technology, Hefei 230026, China}
\affiliation[c]{Shanghai Research Center for Quantum Science and CAS Center for Excellence in Quantum Information and Quantum Physics, University of Science and Technology of China, Shanghai 201315, China}

\emailAdd{dwwu16@fudan.edu.cn}
\emailAdd{tangshanchang@sina.com}
\emailAdd{yu\_shi@ustc.edu.cn}

\abstract{The behavior of accelerating Unruh-DeWitt detectors coupled with a spinor field in (3+1)-dimensional spacetime is investigated. For a single point-like detector with Gaussian switching function, the transition probability increases with the acceleration and thus the  antiUnruh effect   effect cannot occur. Due to the spinor structure of the Dirac field, UV divergences are encountered in the calculation of the entanglement between the detectors. After introducing some UV cutoff $\Lambda$, the logarithmic negativity of detectors is shown to behave nonmonotonically with respect to the acceleration. Besides, the logarithmic negativity increases with the cutoff $\Lambda$ and decreases with the distance between the detectors. The mutual information between the two detectors is also discussed.}

\begin{document}
\maketitle
\flushbottom

\section{Introduction}

Quantum field theory in curved spacetime (QFTCS)~\cite{Wald:1995yp,DEWITT1975295,
Fulling:1989nb,Buchbinder:1992rb,Parker:2009uva} deals with the behavior of quantum fields in presence of classical gravitational fields. Despite not being a fundamental theory, it provides deep insights to what a theory of quantum gravity should look like in low energy. The most important implications in QFTCS are Hawking effect~\cite{Hawking:1975vcx} and Unruh effect~\cite{PhysRevD.14.870,Davies:1974th,PhysRevD.7.2850}. It is well known that quantum entanglement plays a key role in Unruh and Hawking effects. On the other hand, Unruh-DeWitt detectors, as important tools in the study of QFTCS, can be regarded as  qubits in  perspective of  quantum information theory. As a result, the concepts and methods of quantum information can be naturally extended to the study in QFTCS, giving rise to a new research area named relativistic quantum information (RQI).

The study of RQI~\cite{Peres:2002wx,Mann_2012} investigates how concepts in quantum information theory are affected by relativistic motions or gravitational fields. In particular, there are numerous studies on accelerating Unruh-DeWitt detectors coupled with a scalar field. The behavior of a single detector has been carefully studied~\cite{PhysRevD.29.1047} and the decoherence of Unruh-DeWitt detectors due to acceleration was noted~\cite{Kok:2003mc}. For two Unruh-DeWitt detectors, people are interested in their entanglement properties. It has been  discovered that the entanglement is degraded due to acceleration~\cite{Dai:2015ota,Dai:2015dqt}. Detectors in separable states remain separable if they interact with different scalar fields~\cite{Dai:2015ota,Dai:2015dqt}. However, separable detectors could be entangled if they interact with the same scalar field. Such a process arises from the entanglement structure of quantum fields and is called ``entanglement harvesting''~\cite{VALENTINI1991321,Reznik:2002fz,Reznik:2003mnx,PhysRevD.79.044027,Lin:2010zzb,Salton:2014jaa}. There are few studies, however, on the correlations of Unruh-DeWitt detectors coupled with a Dirac field. These questions are conceptually important~\cite{Alsing:2006cj,Fuentes:2010dt,Martin-Martinez:2010bcj,Montero:2011ai,Montero:2011sx} but hard to deal with, because ultraviolet divergences are encountered even for the problem of one detector~\cite{Hummer:2015xaa}. At first it was expected that the divergences are mainly due to the detector quadratically coupled to the field, with the interaction   given as $\mu \bar{\Psi}\Psi$ in the Takagi model~\cite{Takagi:1986kn}, rather than the statistics of the quantum fields~\cite{PhysRevD.96.085012}.  Recently~\cite{Wu:2023glc}, however, it was shown that the entanglement between detectors quadratically coupled with a scalar field is finite. In ~\cite{Wu:2023glc}, the perspective of Rindler rather than inertial observers is used. Moreover, the entanglement depends on the acceleration nonmonotonically. Therefore, given the results for quadratic coupling with scalar fields, it is natural to  extend the methods in to the case of Dirac fields.

In this paper, we study accelerating Unruh-DeWitt detectors coupled with a  spinor field, by  using the methods developed in ~\cite{Wu:2023glc}. We show that despite the finiteness found in ~\cite{Wu:2023glc}, ultraviolet divergences are still encountered for detectors  quadratically coupled with a Dirac field. However, our calculation explicitly shows that such divergences come from the statistical property of the spinor field. In order to study the dependence of entanglement on the acceleration, we introduce a UV cutoff $\Lambda$ in the frequency of the Rindler modes. The nonmonotonicity of entanglement discovered in ~\cite{Wu:2023glc} is also possible in the case of spinor fields if we choose a relatively large cutoff $\Lambda$. Such nonmonotonicity vanishes as the cutoff $\Lambda$ decreases. Besides, the entanglement between the detectors decreases with the increase of the distance. We also investigate the mutual information between the two detectors. It is not dependent on the divergences in the density matrix, and increases with the acceleration monotonically. In addition, our calculation immediately gives the results for a single detector. We find that the transition probability increases with the acceleration and thus there is no antiUnruh effect   effect~\cite{Brenna:2015fga}.

This paper is organized as follows. In Section \ref{sec:Model} we briefly review the Takagi model~\cite{Takagi:1986kn} for the interaction between the Unruh-DeWitt detectors and the Dirac field. Then the reduced density matrix for the final state of each  detector is obtained. In Section \ref{sec:L}, we calculate the convergent terms in the density matrix. These terms reduce to the transition probability for a single detector and the existence of antiUnruh effect is discussed. In Section \ref{sec:M}, we calculate the divergent terms in the density matrix. We introduce a UV cutoff to discuss the logarithmic negativity between the detectors. Section \ref{Sec:con} is the conclusion and outlook.

\section{Model}
\label{sec:Model}
We consider two accelerating two-level Unruh-DeWitt detectors in (3+1)-dimensional spacetime whose trajectories are given as
\begin{equation}
    t=a^{-1}e^{a\xi}\sinh{a\tau},\;\; z=a^{-1}e^{a\xi}\cosh{a\tau}
\end{equation}
where $\xi$ fixed and $\tau$ is the proper time of the detectors. It is considered here that these two detectors share the same acceleration and coincide in the accelerating direction $z$. The coordinate frame is set to be such that they also coincide in one of the remaining direction $y$ and are separated with  distance $x_0$ along direction $x$. 
The interaction between the detectors and the spinor field is described by the Takagi model~\cite{Takagi:1986kn} and the Hamiltonian is
\begin{equation}
  H_2=H_A+H_B,
\end{equation}
and for $j=a,b,$ 
\begin{equation}
H_j=\lambda_j\chi_j(\tau)
\int_{\Sigma_j}N\left[\bar{\Psi}\Psi\right]
\left[f_j(\vec{x})\sigma_j^+e^{i\Omega_j\tau}
+f_j(\vec{x})^*\sigma_j^-e^{-i\Omega_j\tau}\right]
\sqrt{-g}d^3\vec{x},
\end{equation}
where $\lambda$ is the coupling constant; $\chi$ is the switching function; $f$ is the shape function; $\sigma^+$ and $\sigma^-$ are the raising and lowering operators of the two-level system and $\Omega$ is the energy gap. We introduce the normal ordering of the field operators $N\left[\bar{\Psi}\Psi\right]$ to ensure its vacuum expectation value to be zero, so that unnecessary divergences can be avoided. For simplicity, we assume the initial state of the detector to be the ground state $|G\rangle$ and the field is in Minkowski vacuum state $|0_M\rangle$. Then following the standard perturbation procedure as in ~\cite{PhysRevD.96.085012}, we obtain the final reduced density of the detectors. We have used  the fact that the vacuum expectation value  of the field operator $N\left[\bar{\Psi}\Psi\right]$ is zero. 
\begin{equation}
  \rho_r^\Psi=\begin{pmatrix}
      1-P_A^\Psi-P_B^\Psi &   0   &   0   &   M^{\Psi*}\\
      0         &   P_A^\Psi &   L_{AB}^{\Psi*} & 0\\
      0         &   L_{AB}^\Psi & P_B^\Psi   & 0\\
      M^\Psi         &   0   &   0   &   0
    \end{pmatrix}+o(\lambda^2),
\end{equation}
and
\begin{align}
 P_j^\Psi =&\lambda_j^2\int d\tau_1\int d\tau_2  \chi_{j1} \chi_{j2}e^{i\Omega_j(\tau_2-\tau_1)}\int_{\Sigma_{j2}}{d}^3\vec{x}_2f_{j2}\int_{\Sigma_{j1}}{d}^3\vec{x}_1f_{j1}^*\notag\\
    &\times \langle0_M|N\left(\bar{\Psi}_1\Psi_1\right)N\left(\bar{\Psi}_2\Psi_2\right)|0_M\rangle,\\
  L_{AB}^\Psi =&\lambda_A\lambda_B\int d\tau_1\int d\tau_2  \chi_{A1} \chi_{B2} e^{-i\Omega_A\tau_1+i\Omega_B\tau_2}\int_{\Sigma_{A1}}{d}^3\vec{x}_1f_{A1}^*\int_{\Sigma_{B2}}{d}^3\vec{x}_2f_{B2} \notag\\
  &\times \langle0_M|N\left(\bar{\Psi}_1\Psi_1\right)N\left(\bar{\Psi}_2\Psi_2\right)|0_M\rangle,\label{eq:(2,6)}\\
  M^\Psi =&\lambda_A\lambda_B\int d\tau_1\int d\tau_2\chi_{A1}\chi_{B2}e^{i\Omega_A\tau_1+i\Omega_B\tau_2}\int_{\Sigma_{A1}}f_{A1}{d}^3\vec{x}_1\int_{\Sigma_{B2}}f_{B2}{d}^3\vec{x}_2 \notag\\
  &\times \langle 0_M|\mathcal{T}\left[N\left(\bar{\Psi}_1\Psi_1\right)N\left(\bar{\Psi}_2\Psi_2\right)\right]|0_M\rangle.\label{eq:(2,7)}
\end{align}
Sometimes we omit the determinant of the metric tensor, as later we would assume $\xi=0$ in our calculation. 

The entries of the reduce density matrix take the form of the Fourier transformation of the time-ordered Wightman function of the field operator $\langle0_M|N\left(\bar{\Psi}_1\Psi_1\right)
N\left(\bar{\Psi}_2\Psi_2\right)|0_M\rangle$. It is well known that the propagators of fields with nonzero spins can be expressed in terms of  that of a  field  with spin-$0$~\cite{Mandl:1985bg,Weinberg:1995mt}. Therefore we can reduce the Wightman function of $N\left[\bar{\Psi}\Psi\right]$ by (the proof can be found in Appendix \ref{App:Correlation function})
\begin{align}  \notag
V_\Psi=&\langle0_M|N\left(\bar{\Psi}_1
\Psi_1\right)N\left(\bar{\Psi}_2\Psi_2\right)|0_M\rangle\\ \notag
=&-4\eta^{\mu\nu}\left(\partial_\mu\langle0_M|\hat{\Phi}_1\hat{\Phi}_2|0_M\rangle\right)\left(\partial_\nu\langle0_M|\hat{\Phi}_1\hat{\Phi}_2|0_M\rangle\right)-4m^2\langle0_M|\hat{\Phi}_1\hat{\Phi}_2|0_M\rangle^2.\\ \notag
=&-4e^{-2a\xi}\left(\partial_\tau\langle0_M|\hat{\Phi}_1\hat{\Phi}_2|0_M\rangle\right)^2 +4e^{-2a\xi}\left(\partial_\xi\langle0_M|\hat{\Phi}_1\hat{\Phi}_2|0_M\rangle\right)^2 \\ \notag
    &+4\left(\nabla_{\vec{x}_\bot}\langle0_M|\hat{\Phi}_1\hat{\Phi}_2|0_M\rangle\right)^2 -4m^2\langle0_M|\hat{\Phi}_1\hat{\Phi}_2|0_M\rangle^2\\
    \equiv& -4e^{-2a\xi}V_\Psi^\tau+4e^{-2a\xi}V_\Psi^\xi+4V_\Psi^\bot-4m^2V_\Psi^m. \label{eq:2.8}
\end{align}
In the third step we have made of a transformaiton of variables. The Wightman function of the scalar field in terms of Rindler modes is given as~\cite{Wu:2023glc}
\begin{align} \notag
    &\langle 0_M|\hat{\Phi}(\tau_1,\vec{x}_1)\hat{\Phi}(\tau_2,\vec{x}_2)|0_M\rangle\\
    =&\int_0^\infty d\omega\iint d^2\vec{k}_{\bot} \left(v^R_{1\omega\vec{k}_{\bot}}v^{R*}_{2\omega\vec{k}_{\bot}}
+v^{R*}_{1\omega\vec{k}_{\bot}}v^R_{2\omega\vec{k}_{\bot}}e^{-2\pi\omega/a}\right)\frac{1}{1-e^{-2\pi\omega/a}},\label{eq:2.9}
\end{align}
where $ v^R_{\omega\vec{k}_\bot}$ is the Rindler mode~\cite{Crispino:2007eb}
\begin{equation}
  v^R_{\omega\vec{k}_\bot}=\left[\frac{\sinh(\pi\omega/a)}{4\pi^4a}\right]^{1/2}K_{i\omega/a}\left(\frac{\kappa}{a}e^{a\xi}\right)e^{i\vec{k}_\bot\cdot\vec{x}_\bot-i\omega\tau}.\label{eq:2.10}
\end{equation}
Therefore we are able to extend the method in ~\cite{Wu:2023glc} to the case of Dirac fields. The calculation techniques in this paper is quite similar to that in ~\cite{Wu:2023glc}.

\section{Calculation of $L$ and the  antiUnruh effect   effect}
\label{sec:L}
In this section we are to calculate $L$ by using the method developed in ~\cite{Wu:2023glc}. Calculation of $P$ directly follows. It can be easily checked that $P$ is the transition probability of a single Unruh-DeWitt detector coupled with a Dirac field~\cite{Gray:2018ifq}. As a result, the calculation of $L$ allows us to discuss the existence of antiUnruh effect for spinor fields in (3+1)-dimensional spacetime.

According to Eq.~\eqref{eq:(2,6)} and Eq.~\eqref{eq:2.9}, we divide $L_{AB}^\Psi$ as 
\begin{align}
  L_{AB}^\Psi=&4\lambda_A\lambda_B\left(-L^{\Psi\tau}_{AB}+L^{\Psi\xi}_{AB}+L^{\Psi\vec{x}_\bot}_{AB}\right)-2m^2\lambda_A\lambda_BL^{\Phi^2}_{AB},\label{eq:3.1}
\end{align}
where $L^{\Phi^2}_{AB}$ is the $L$ term for quadratic coupling with scalar fields and has  already been treated in ~\cite{Wu:2023glc}.
Later we will assume the Dirac field to be massless so this term is somewhat irrelevant in our final calculations. We write the forms of other terms as following. For $i=\tau$,$\xi$ and $\vec{x}_\bot$,
\begin{align}
 L^{\Psi i}_{AB}\label{eq:3.2}
    =&\int_0^\infty d\omega_1\int_0^\infty d\omega_2\iint d^2\vec{k}_{1\bot}
    \iint d^2\vec{k}_{2\bot} \frac{1}{1-e^{-2\pi\omega_1/a}}\frac{1}{1-e^{-2\pi\omega_2/a}}\\\notag &\times\left[I^{i0}_{\omega_1\vec{k}_{1\bot},\omega_2\vec{k}_{2\bot}} +I^{i2}_{\omega_1\vec{k}_{1\bot},\omega_2\vec{k}_{2\bot}}e^{-2\pi\omega_2/a} +I^{i1}_{\omega_1\vec{k}_{1\bot},\omega_2\vec{k}_{2\bot}}e^{-2\pi\omega_1/a} +I^{i12}_{\omega_1\vec{k}_{1\bot},\omega_2\vec{k}_{2\bot}}e^{-2\pi(\omega_1+\omega_2)/a}\right],
\end{align}
where the $I$ terms can be calculated as in ~\cite{Wu:2023glc} by means of the Rindler modes and the Fourier transformation of switching functions, for instance,
\begin{align}\notag
I^{\tau0}_{\omega_1\vec{k}_{1\bot},\omega_2\vec{k}_{2\bot}}
  =&\int d\tau_1\int d\tau_2  \chi_{A1} \chi_{B2} e^{-i\Omega_A\tau_1+i\Omega_B\tau_2}\int_{\Sigma_{A1}}e^{-2a\xi_1} {d}^3\vec{x}_1f_{A1}^*\int_{\Sigma_{B2}} {d}^3\vec{x}_2f_{B2}\\\notag
  &\times \left(\partial_{\tau_1}v^R_{1\omega_1\vec{k}_{1\bot}}\right)\left(\partial_{\tau_1}v^R_{1\omega_2\vec{k}_{2\bot}}\right)v^{R*}_{2\omega_1\vec{k}_{1\bot}}v^{R*}_{2\omega_2\vec{k}_{2\bot}}\\\notag
  =&-\omega_1\omega_2\int d\tau_1\int d\tau_2  \chi_{A1} \chi_{B2} e^{-i\Omega_A\tau_1+i\Omega_B\tau_2}\int_{\Sigma_{A1}}e^{-2a\xi_1} {d}^3\vec{x}_1f_{A1}^*\int_{\Sigma_{B2}} {d}^3\vec{x}_2f_{B2}\\\notag
  &\times v^R_{1\omega_1\vec{k}_{1\bot}}v^R_{1\omega_2\vec{k}_{2\bot}}v^{R*}_{2\omega_1\vec{k}_{1\bot}}v^{R*}_{2\omega_2\vec{k}_{2\bot}}\\
  =&-\omega_1\omega_2\eta^{0A'*}_{\omega_1\vec{k}_{1\bot},\omega_2\vec{k}_{2\bot}}\eta^{0B}_{\omega_1\vec{k}_{1\bot},\omega_2\vec{k}_{2\bot}}.\label{eq:3.3}
\end{align}
The $\eta$ terms in Eq.~\eqref{eq:3.3} are the same as those for quadratic coupling with a scalar field in ~\cite{Wu:2023glc}. Therefore, compared to the case of quadratic  coupling with a scalar field, we have extra factors $\omega_1 \omega_2$ due to the derivative of the Rindler modes for the case of coupling with a Dirac field. The same results apply to $I^{\vec{x}_{\bot}}$ terms, with the extra factors being $\vec{k}_{1\bot}\cdot\vec{k}_{2\bot}$. The details are given in the Appendix \ref{App: Calculation of $L$}.

The Rindler modes depend on $\xi$ in the form of modified Bessel function in Eq.~\eqref{eq:2.10}. The derivation  is given in Appendix \ref{App:Special function and integrals}. Therefore, 
\begin{align}\notag
I^{\xi0}_{\omega_1\vec{k}_{1\bot},\omega_2\vec{k}_{2\bot}}
  =&2\pi \frac{\sinh(\pi\omega_1/a)\sinh(\pi\omega_2/a)}{4(4\pi^4a)^2} G_B(\Omega_B+\omega_1+\omega_2)G_A(-\Omega_A-\omega_1-\omega_2) \\ \notag&\times\int_{\Sigma_{A1}} {d}^3\vec{x}_1f_{A1}^*\int_{\Sigma_{B2}} {d}^3\vec{x}_2f_{B2}
  e^{i(\vec{k}_{1\bot}+\vec{k}_{2\bot})\cdot(\vec{x}_{1\bot}-\vec{x}_{2\bot})}\\\notag
  &\hspace{-3cm}\times\kappa_1\kappa_2\left[K_{i\omega_1/a-1}\left(\frac{\kappa_1}{a}e^{a\xi_1}\right) +K_{i\omega_1/a+1}\left(\frac{\kappa_1}{a}e^{a\xi_1}\right)\right] \left[K_{i\omega_2/a-1}\left(\frac{\kappa_2}{a}e^{a\xi_1}\right) +K_{i\omega_2/a+1}\left(\frac{\kappa_2}{a}e^{a\xi_1}\right)\right] \\
  &\times K_{i\omega_1/a}\left(\frac{\kappa_1}{a}e^{a\xi_2}\right) K_{i\omega_2/a}\left(\frac{\kappa_2}{a}e^{a\xi_2}
  \right),
\end{align}
where $G_A$ and $G_B$ are the Fourier transforms of the switching function $\chi$. The other terms are also given in Appendix \ref{App: Calculation of $L$}.

\subsection{Point-like and Gaussian detectors: antiUnruh effect   effect}
As in ~\cite{Wu:2023glc}, we assume the detectors to be point-like and the switching function is Gaussian, i.e.
\begin{align}
  &\chi_A(\tau)=\frac{1}{\sqrt{2\pi}\sigma_A}\exp\left(-\frac{\tau^2}{2\sigma_A^2}\right),\;\;f_A(\xi,x,y)=\delta(\xi)\delta(x)\delta(y),\\
  &\chi_B(\tau)=\frac{1}{\sqrt{2\pi}\sigma_B}\exp\left[-\frac{(\tau-\tau_0)^2}{2\sigma_B^2}\right],\;\;f_B(\xi,x,y)=\delta(\xi)\delta(x-x_0)\delta(y).
\end{align}
The Fourier transform of the switching function is given explicitly,
\begin{align}
  &G_A(\nu)=\exp\left(-\frac{\sigma_A^2}{2}\nu^2\right),\\
  &G_B(\nu)=\exp\left(-\frac{\sigma_B^2}{2}\nu^2+i\nu\tau_0\right).
\end{align}
With such assumptions the space integral is immediately carried out and in order to calculate $L_{AB}^\Psi$ we are left with the integrals over $\omega$ and $\vec{k}_\bot$. Due to the symmetry in $x-y$ plane, the integral over $\vec{k}_\bot$ can be done using polar coordinates (the Dirac field is considered to be massless in this step). The calculation and results are given in Appendix \ref{App: Calculation of $L$}. 

\begin{figure}[htb!]
  \centering
  \includegraphics[width=0.8\textwidth]{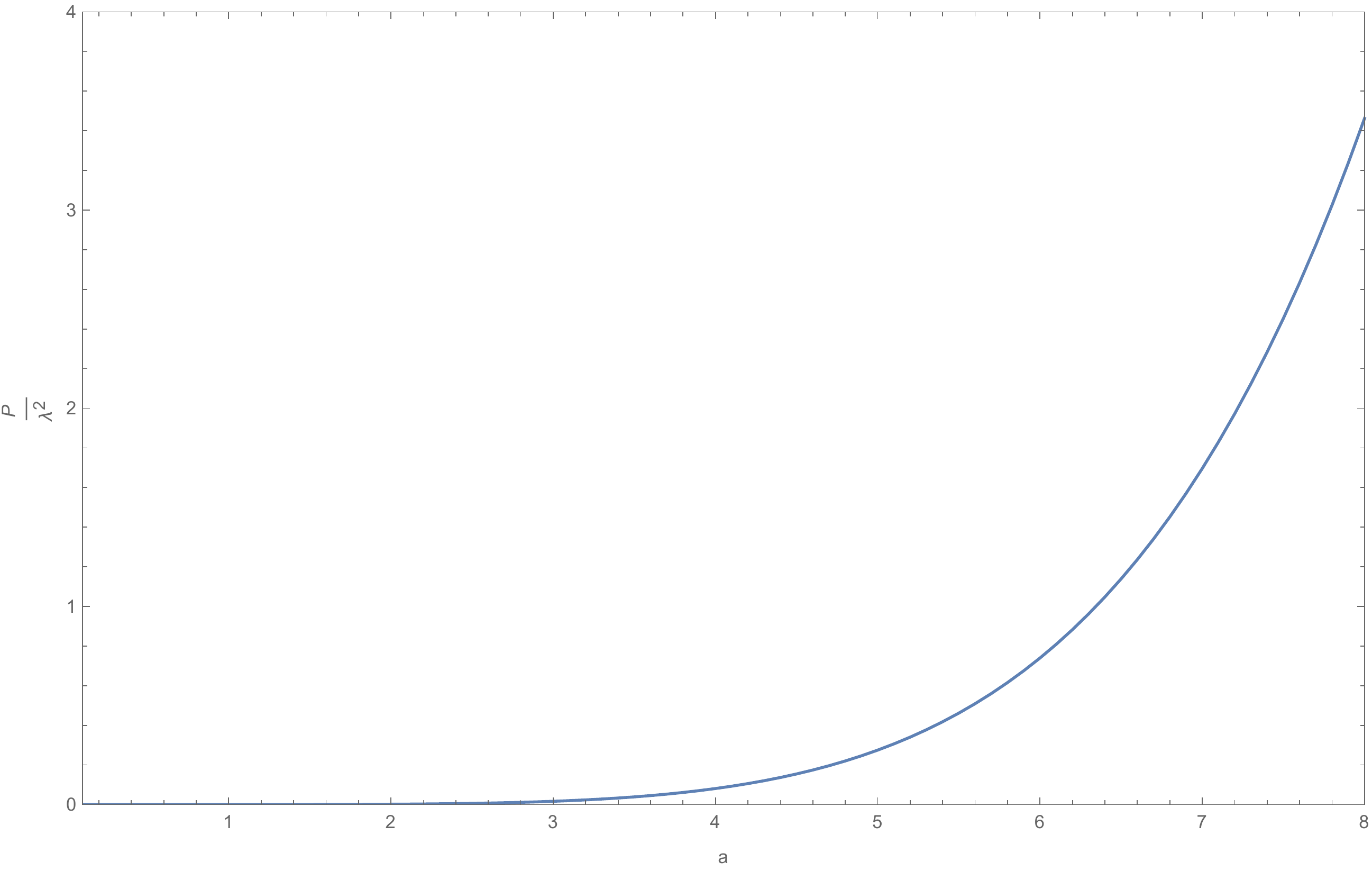}
  \caption[Transition probability of a detector coupled with a Dirac field ($\sigma=1,\Omega=1$)]{%
Transition probability of a detector coupled with a Dirac field ($\sigma=1,\Omega=1$)%
   }
   \label{fig:DiracP}
\end{figure}

Here we further require the two detectors to be identical ($\tau_0=0$, $x_0=0$) and $L$ terms reduce to the transition probability $P$ as
\begin{align}
  \begin{split}
    P_{AB}^{\Psi\tau}=&\frac{1}{32\pi^3}\int_0^\infty d\omega_1\int_0^\infty d\omega_2\frac{\omega_1^2\omega_2^2}{\sinh(\pi\omega_1/a)\sinh(\pi\omega_2/a)}\\
    &\hspace{-2cm}\times\left\{-\exp\left[-\sigma^2(\omega_1+\omega_2+\Omega)^2+\pi(\omega_1+\omega_2)/a\right] +\exp\left[-\sigma^2(\Omega+\omega_1-\omega_2)^2+\pi(\omega_1-\omega_2)/a\right]\right.\\ &\hspace{-2cm}\left.+\exp\left[-\sigma^2(\Omega-\omega_1+\omega_2)^2-\pi(\omega_1-\omega_2)/a\right]
    -\exp\left[-\sigma^2(\Omega-\omega_1-\omega_2)^2-\pi(\omega_1+\omega_2)/a\right]\right\},
  \end{split}\\
  \begin{split}
    P_{AB}^{\Psi\xi}=&\frac{a^2}{32\pi^3}\int_0^\infty d\omega_1\int_0^\infty d\omega_2\frac{\omega_1\omega_2}{\sinh(\pi\omega_1/a)\sinh(\pi\omega_2/a)}\\
    &\hspace{-2cm}\times\left\{\exp\left[-\sigma^2(\omega_1+\omega_2+\Omega)^2+\pi(\omega_1+\omega_2)/a\right] +\exp\left[-\sigma^2(\Omega+\omega_1-\omega_2)^2+\pi(\omega_1-\omega_2)/a\right]\right.\\ &\hspace{-2cm}\left.+\exp\left[-\sigma^2(\Omega-\omega_1+\omega_2)^2-\pi(\omega_1-\omega_2)/a\right]
    +\exp\left[-\sigma^2(\Omega-\omega_1-\omega_2)^2-\pi(\omega_1+\omega_2)/a\right]\right\},
  \end{split}\\
  \begin{split}
    P^{\Psi\vec{x}_\bot}_{AB}=0.
  \end{split}
\end{align}
The integral over $\omega_1$ and $\omega_2$ can be done numerically  using Mathematica, and we draw the relation between transition probability and the acceleration in figure~\ref{fig:DiracP}.

After some attempts on different parameters, one can observe that the transition probability increases with the acceleration monotonically, which implies that there is no antiUnruh effect for point-like and Gaussian Unruh-DeWitt detectors which are coupled with a Dirac field in (3+1)-dimensional spacetime. Same results hold for detectors coupled with a scalar field linearly or quadratically~\cite{Wu:2023glc}. On the other hand, antiUnruheffect is found for the same detectors but coupled with scalar fields in (1+1)-dimensional spacetime~\cite{Brenna:2015fga}. Therefore, our discussion seems to indicate the increase of the spacetime dimensions suppresses the antiUnruh effect , (this is in contrast with some other studies such as ~\cite{Yan:2022xgg} which claims that the Unruh effect in small spacetime dimensions is able to become the antiUnruh effect in large spacetime dimensions).

\section{Calculation of $M$ and the correlations between Unruh-DeWitt detectors}
\label{sec:M}

The calculation of $M$ involves the ``time-ordered Wightman function" which can be written as
\begin{align}
\notag
    &\langle 0_M|\mathcal{T}\left[N\left(\bar{\Psi}_1\Psi_1\right)N\left(\bar{\Psi}_2\Psi_2\right)\right]|0_M\rangle\\\notag
    =&\theta(\tau_1-\tau_2)\Big[-4e^{-2a\xi_1}\left(\partial_\tau\langle0_M|\hat{\Phi}_1\hat{\Phi}_2|0_M\rangle\right)^2 +4e^{-2a\xi_1}\left(\partial_\xi\langle0_M|\hat{\Phi}_1\hat{\Phi}_2|0_M\rangle\right)^2 \\\notag
&+4\left(\nabla_{\vec{x}_\bot}\langle0_M|\hat{\Phi}_1\hat{\Phi}_2|0_M\rangle\right)^2 -4m^2\langle0_M|\hat{\Phi}_1\hat{\Phi}_2|0_M\rangle^2\Big]\\\notag
    +&\theta(\tau_2-\tau_1)\Big[-4e^{-2a\xi_2}\left(\partial_\tau\langle0_M|\hat{\Phi}_2\hat{\Phi}_1|0_M\rangle\right)^2 +4e^{-2a\xi_2}\left(\partial_\xi\langle0_M|\hat{\Phi}_2\hat{\Phi}_1|0_M\rangle\right)^2 \\\notag
&+4\left(\nabla_{\vec{x}_\bot}\langle0_M|\hat{\Phi}_2\hat{\Phi}_1|0_M\rangle\right)^2 -4m^2\langle0_M|\hat{\Phi}_2\hat{\Phi}_1|0_M\rangle^2\Big]\\
    \equiv& -4D_\Psi^\tau+4D_\Psi^\xi+4D_\Psi^\bot-4m^2D_\Psi^m. \label{eq:4.1}
\end{align}

The explicit form of $D$s directly follows from Eq.~\eqref{eq:2.9} and is given in Appendix \ref{App: Calculation of $M$}. Substitute Eq.~\eqref{eq:4.1} into Eq.~\eqref{eq:(2,7)}, and we can also divide $M^\Psi$ into
\begin{equation}
    M^\Psi=4\lambda_A\lambda_B\left(-M^{\Psi\tau}+M^{\Psi\xi}+M^{\Psi\vec{x}_\bot}\right)-2m^2\lambda_A\lambda_B M^{\Phi^2},\label{eq:4.2}
\end{equation}
where $M^{\Psi i}$ is of the form (for $i=\tau,\xi,\vec{x}_\bot$)
\begin{align}
M^{\Psi i}
    =&\int_0^\infty d\omega_1\int_0^\infty d\omega_2\iint d^2\vec{k}_{1\bot}
    \iint d^2\vec{k}_{2\bot} \frac{1}{1-e^{-2\pi\omega_1/a}}\frac{1}{1-e^{-2\pi\omega_2/a}}\\\notag
&\times\left[W^{i0}_{\omega_1\vec{k}_{1\bot},\omega_2\vec{k}_{2\bot}} +W^{i2}_{\omega_1\vec{k}_{1\bot},\omega_2\vec{k}_{2\bot}}e^{-2\pi\omega_2/a} +W^{i1}_{\omega_1\vec{k}_{1\bot},\omega_2\vec{k}_{2\bot}}e^{-2\pi\omega_1/a} +W^{i12}_{\omega_1\vec{k}_{1\bot},\omega_2\vec{k}_{2\bot}}e^{-2\pi(\omega_1+\omega_2)/a}\right].\label{eq:4.3}
\end{align}
The meaning of $W$s is similar to that of $I$s in Eq.~\eqref{eq:3.2}: $W^i$s correspond to terms arising from the 'i' derivative of the Wightman functions. For example, 
\begin{align} \notag
W^{\tau0}_{\omega_1\vec{k}_{1\bot},\omega_2\vec{k}_{2\bot}}
  =&\int d\tau_1\int d\tau_2\chi_{A1}\chi_{B2}e^{i\Omega_A\tau_1+i\Omega_B\tau_2}\int_{\Sigma_{A1}}f_{A1} {d}^3\vec{x}_1\int_{\Sigma_{B2}}f_{B2} {d}^3\vec{x}_2\\\notag
  &\times\Bigg[\theta(\tau_1-\tau_2)e^{-2a\xi_1}\left(\partial_{\tau_1}v^R_{1\omega_1\vec{k}_{1\bot}}\right)\left(\partial_{\tau_1}v^R_{1\omega_2\vec{k}_{2\bot}}\right)v^{R*}_{2\omega_1\vec{k}_{1\bot}}v^{R*}_{2\omega_2\vec{k}_{2\bot}} \\\notag
  &+\theta(\tau_2-\tau_1)e^{-2a\xi_2}\left(\partial_{\tau_2}v^R_{2\omega_1\vec{k}_{1\bot}}\right)\left(\partial_{\tau_2}v^R_{2\omega_2\vec{k}_{2\bot}}\right)v^{R*}_{1\omega_1\vec{k}_{1\bot}}v^{R*}_{1\omega_2\vec{k}_{2\bot}}\Bigg]\\\notag
  =&-\omega_1\omega_2\frac{2\pi\sinh(\pi\omega_1/a)\sinh(\pi\omega_2/a)}{(4\pi^4a)^2}\frac{-2(\omega_1+\omega_2)}{2\pi i}\int du \frac{G_A(\Omega_A+u)G_B(\Omega_B-u)}{u^2-(\omega_1+\omega_2)^2+i\epsilon}\\
    &\times K_{i\omega_1/a}\left(\frac{\kappa_1}{a}\right) K_{i\omega_2/a}\left(\frac{\kappa_2}{a}\right)K_{i\omega_1/a}\left(\frac{\kappa_1}{a}\right) K_{i\omega_2/a}\left(\frac{\kappa_2}{a}\right)e^{-i(\vec{k}_{1\bot}+\vec{k}_{2\bot})\cdot(\vec{x}_{1\bot}-\vec{x}_{2\bot})},
\end{align}
where in the last step we use the same technique in ~\cite{Wu:2023glc}. That is, we replace the Heaviside step function $\theta$ by contour integrals in complex $u$ plane, and then we make use of the positive frequency modes and the Fourier transform of the switching function to integrate $\tau$. Note that our method of contour integral is valid only when $\xi=0$ (so the detectors are point-like and thus the integral over space vanishes). As before, we assume the switching functions of the detectors to be synchronously Gaussian, i.e., the Fourier transform is $G_i(\nu)=\exp\left(-\frac{\sigma_i^2}{2}\nu^2\right)$. Then by the formula~\cite{Wu:2023glc} 
\begin{equation}
  \begin{split} 
    \int_0^{\infty} du\frac{e^{-\sigma^2 u^2}}{u^2-\omega^2+ i\epsilon}
    =\frac{\pi e^{-\sigma^2\omega^2}}{2i\omega}\text{erfc}(i\sigma\omega),
  \end{split}
\end{equation}
we obtain 
\begin{equation}
\begin{split}
W^{\tau0}_{\omega_1\vec{k}_{1\bot},\omega_2\vec{k}_{2\bot}}=&-\omega_1\omega_2\frac{\sinh(\pi\omega_1/a)\sinh(\pi\omega_2/a)}{(4\pi^4a)^2/2\pi}K_{i\omega_1/a}\left(\frac{\kappa_2}{a}\right)^2K_{i\omega_2/a}\left(\frac{\kappa_2}{a}\right)^2 e^{-\sigma^2\Omega^2}\\
&\times e^{i(k_{1x}+k_{2x})x_{0}}e^{-\sigma^2(\omega_1+\omega_2)^2}\text{erfc}\left[i\sigma(\omega_1+\omega_2)\right].
\end{split}
\end{equation}
As in section \ref{sec:L}, compared to the case of quadratic coupling with scalar field, we gain extra factors $\omega_1 \omega_2$ due to the derivative of the Rindler modes and the same results apply to $W^{\vec{x}_{\bot}}$ terms, with the extra factors being $\vec{k}_{1\bot}\cdot\vec{k}_{2\bot}$.

The $W^\xi$ terms can be obtained in similar ways. For example,
\begin{equation}
\begin{split}
W^{\xi0}_{\omega_1\vec{k}_{1\bot},\omega_2\vec{k}_{2\bot}}=&\frac{\kappa_1\kappa_2 \sinh(\pi\omega_1/a)\sinh(\pi\omega_2/a)}{4(4\pi^4a)^2/2\pi}\Big[K_{i\omega_1/a-1}\left(\frac{\kappa_1}{a}\right) +K_{i\omega_1/a+1}\left(\frac{\kappa_1}{a}\right)\Big]\\
&\times \Big[K_{i\omega_2/a-1}\left(\frac{\kappa_2}{a}\right) +K_{i\omega_2/a+1}\left(\frac{\kappa_2}{a}\right)\Big]K_{i\omega_1/a}\left(\frac{\kappa_1}{a}\right) K_{i\omega_2/a}\left(\frac{\kappa_2}{a}\right)\\
&\times e^{-\sigma^2\Omega^2}e^{i(k_{1x}+k_{2x})x_{0}}e^{-\sigma^2(\omega_1+\omega_2)^2}\text{erfc}\left[i\sigma(\omega_1+\omega_2)\right].
\end{split}
\end{equation}
The readers may refer to Appendix \ref{App: Calculation of $M$} for the other $W$ terms.

As in section \ref{sec:L}, the integral over $k_\bot$ can be carried out by polar coordinates and we end up with integrals over $\omega_1$ and $\omega_2$ to get $M^\Psi$ (see Appendix \ref{App: Calculation of $M$} for further details). As is expected, $M^{\Psi_\tau}$ differs from $M^{\Phi^2}$ with factors $\omega_1\omega_2$. 

Unlike the case in which the detectors interact with a scalar field quadratically, the integrals over $\omega_1$ and $\omega_2$ diverge, as we can draw a picture of the imaginary part of the integrand in $M^{\Psi_\tau}$. We see that apart from the fact that the integrand is highly oscillatory, it diverges in the direction of the diagonal of $\omega_1$-$\omega_2$ plane. Our calculations show that the divergence comes from the extra factors such as $\omega_1\omega_2$ in $M^\Psi$.

\begin{figure}[htb!]
  \centering
  \includegraphics[width=1\textwidth]{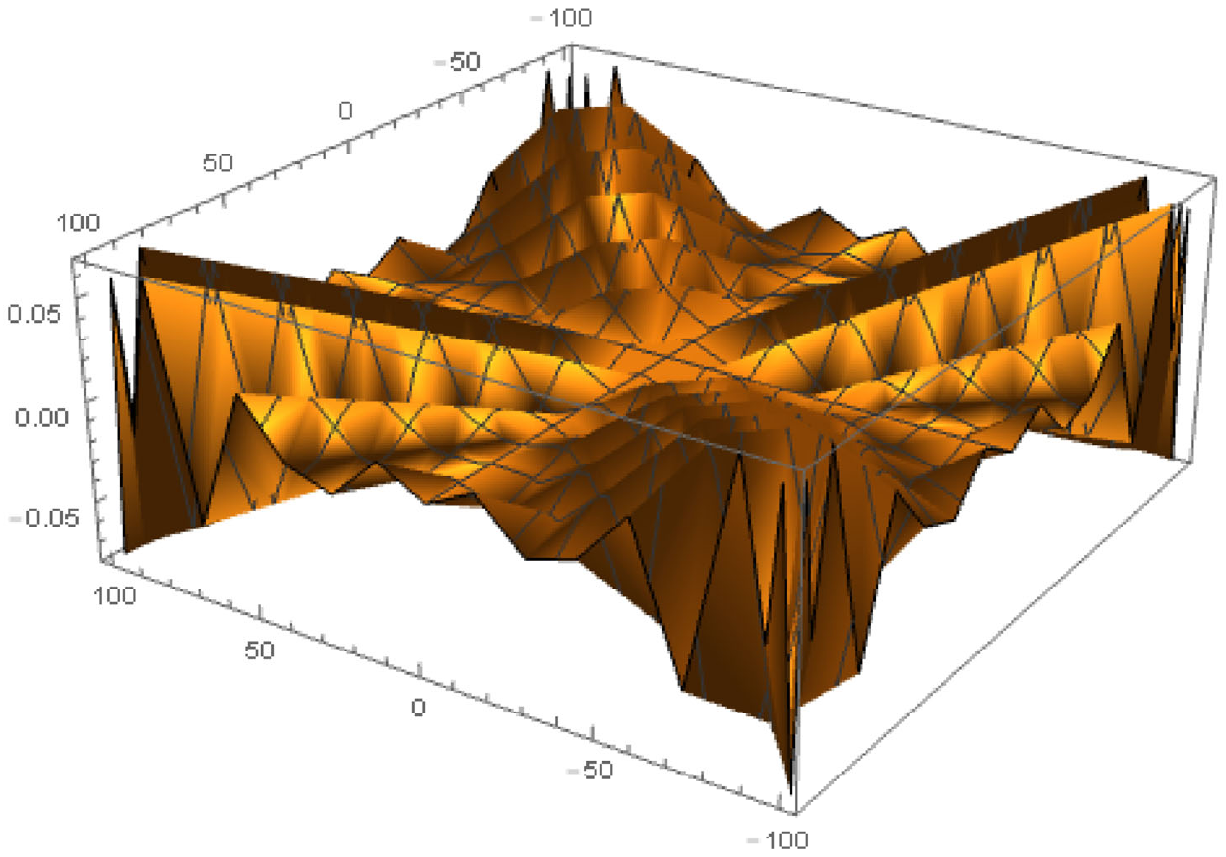}
  \caption[Imaginary part of the integrand in $M^\Psi$ ($\sigma=1,\Omega=1,a=1,x_0=1$)]{%
Imaginary part of the integrand in $M^\Psi$ ($\sigma=1,\Omega=1,a=1,x_0=1$)%
   }
  \label{fig:DiracInt}
\end{figure}

\subsection{Logarithmic negativity}
Despite the divergences encountered in $M$ terms, we can still make some attempts on studying the entanglement between two accelerating Unruh-DeWitt detectors in a spinor field. We introduce a hard UV cutoff to the frequency of the Rindler modes $\omega_1=\omega_2=\Lambda$. The entanglement between the detectors  can be described by logarithmic negativity, which in our case, takes the form~\cite{PhysRevD.96.085012}
\begin{equation}
  \mathcal{N}=\max(|M|-P,0).
\end{equation}

\begin{figure}[htb!]
  \centering
  \includegraphics[width=0.8\textwidth]{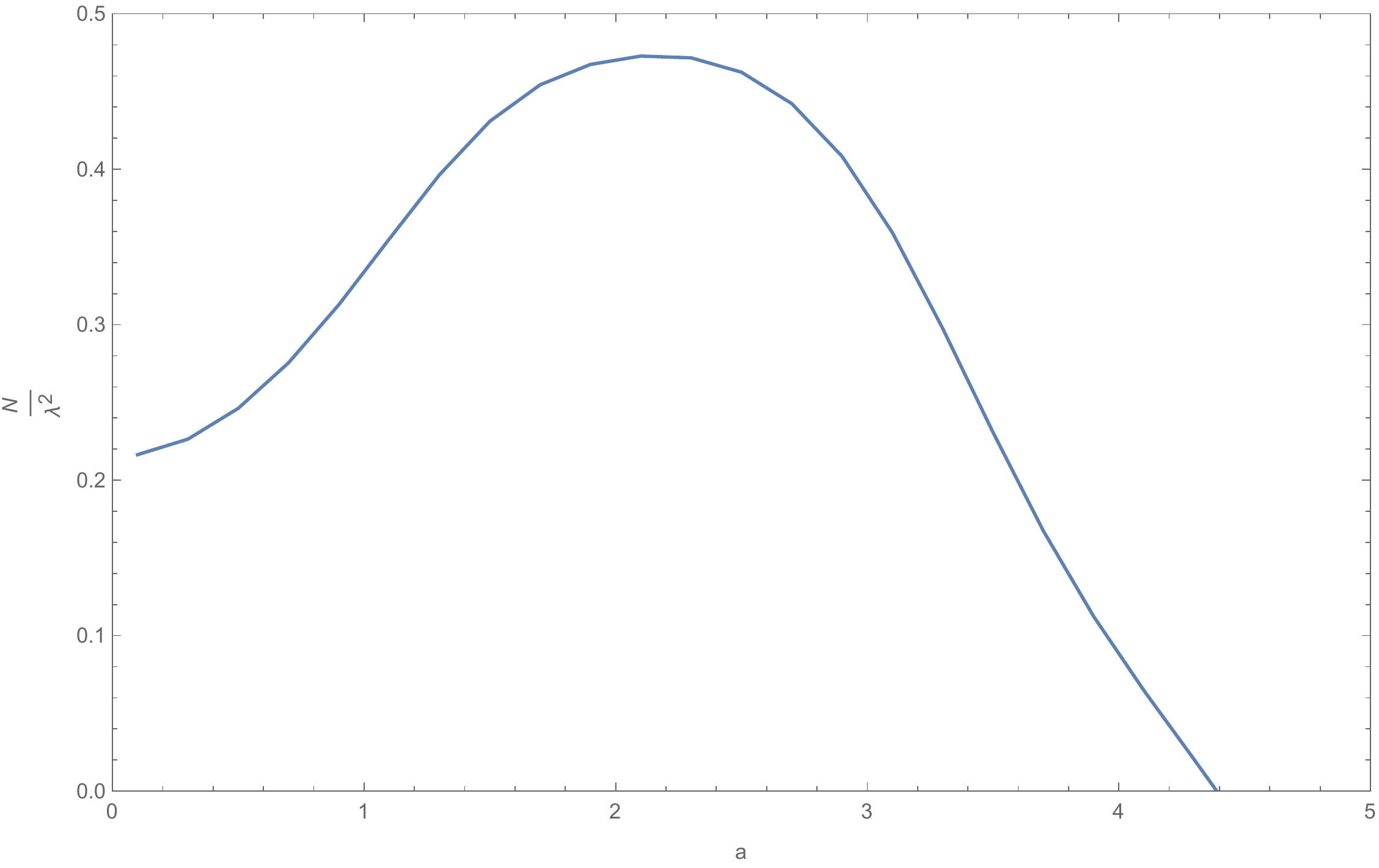}
  \caption[The logarithmic negativity of accelerating Unruh-DeWitt detectors in a Dirac field ($\sigma=1,\Omega=1,x_0=0.7,\Lambda=80$)]{%
The logarithmic negativity of accelerating Unruh-DeWitt detectors in a Dirac field ($\sigma=1,\Omega=1,x_0=0.7,\Lambda=80$)%
   }
  \label{fig:Dirac80}
\end{figure}

Firstly, we choose a relatively large UV cutoff $\Lambda=80$ (unfortunately it might take quite a long time to numerically integrate $M^\Psi$ if the cutoff $\Lambda$ is too large). The dependence of the logarithmic negativity on the acceleration of the detectors is given by figure~\ref{fig:Dirac80}.
It can be seen that the entanglement increases at first and then decreases with the increase of the  acceleration, and at last experiences sudden death at a certain value of acceleration. Such behavior is quite similar to what is obtained  in ~\cite{Wu:2023glc} for detectors coupled quadratically with a scalar field. The nonmonotonicity of logarithmic negativity as a function of the acceleration is also noted elsewhere in different circumstances~\cite{Liu:2021dnl} . Combined with the conclusion in a previous paper~\cite{Wu:2023glc}, it indicates the nonmonotonicity of entanglement to the quadratic form of the interaction Hamiltonian.

\begin{figure}[htb!]
  \centering
  \includegraphics[width=0.8\textwidth]{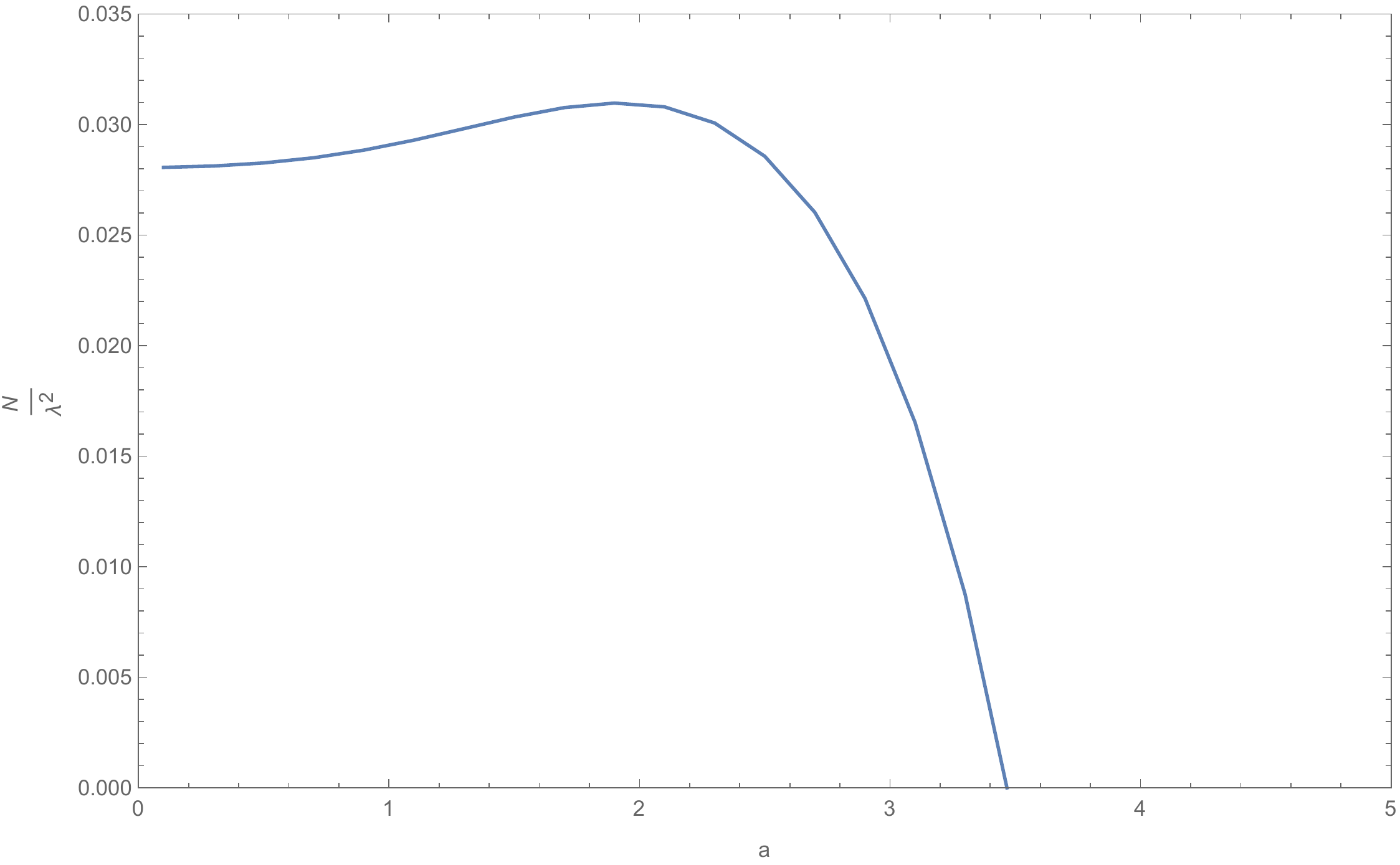}
  \caption[The logarithmic negativity of accelerating Unruh-DeWitt detectors in a Dirac field ($\sigma=1,\Omega=1,x_0=0.7,\Lambda=20$)]{%
The logarithmic negativity of accelerating Unruh-DeWitt detectors in a Dirac field ($\sigma=1,\Omega=1,x_0=0.7,\Lambda=20$)%
   }
  \label{fig:Dirac20-0.7}
\end{figure}

What if we choose a relatively small UV cutoff? It is  shown in figure~\ref{fig:Dirac20-0.7} that the nonmonotonicity of entanglement is still valid  for cutoff $\Lambda=20$.

\begin{figure}[htb!]
  \centering
  \includegraphics[width=0.8\textwidth]{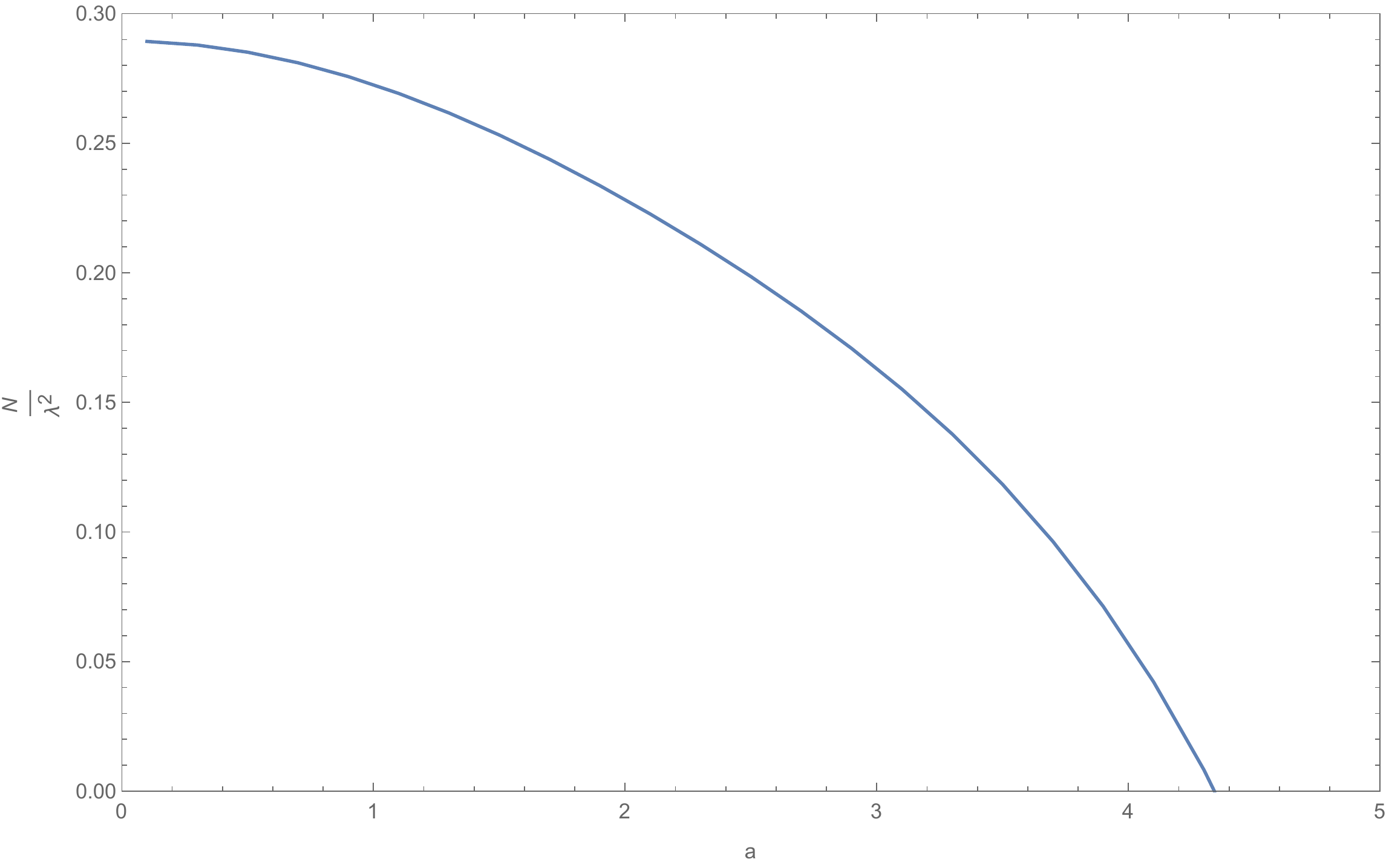}
  \caption[The logarithmic negativity of accelerating Unruh-DeWitt detectors in a Dirac field ($\sigma=1,\Omega=1,x_0=0.5,\Lambda=20$)]{%
The logarithmic negativity of accelerating Unruh-DeWitt detectors in a Dirac field ($\sigma=1,\Omega=1,x_0=0.5,\Lambda=20$)%
   }
  \label{fig:Dirac20-0.5}
\end{figure}

The nonmonotonicity of entanglement could disappear if the distance between the detectors is small enough, as shown in figure~\ref{fig:Dirac20-0.5} ($x_0=0.5$).
It is also natural that the logarithmic negativity increases with the cutoff.

\begin{figure}[htb!]
  \centering
  \includegraphics[width=0.8\textwidth]{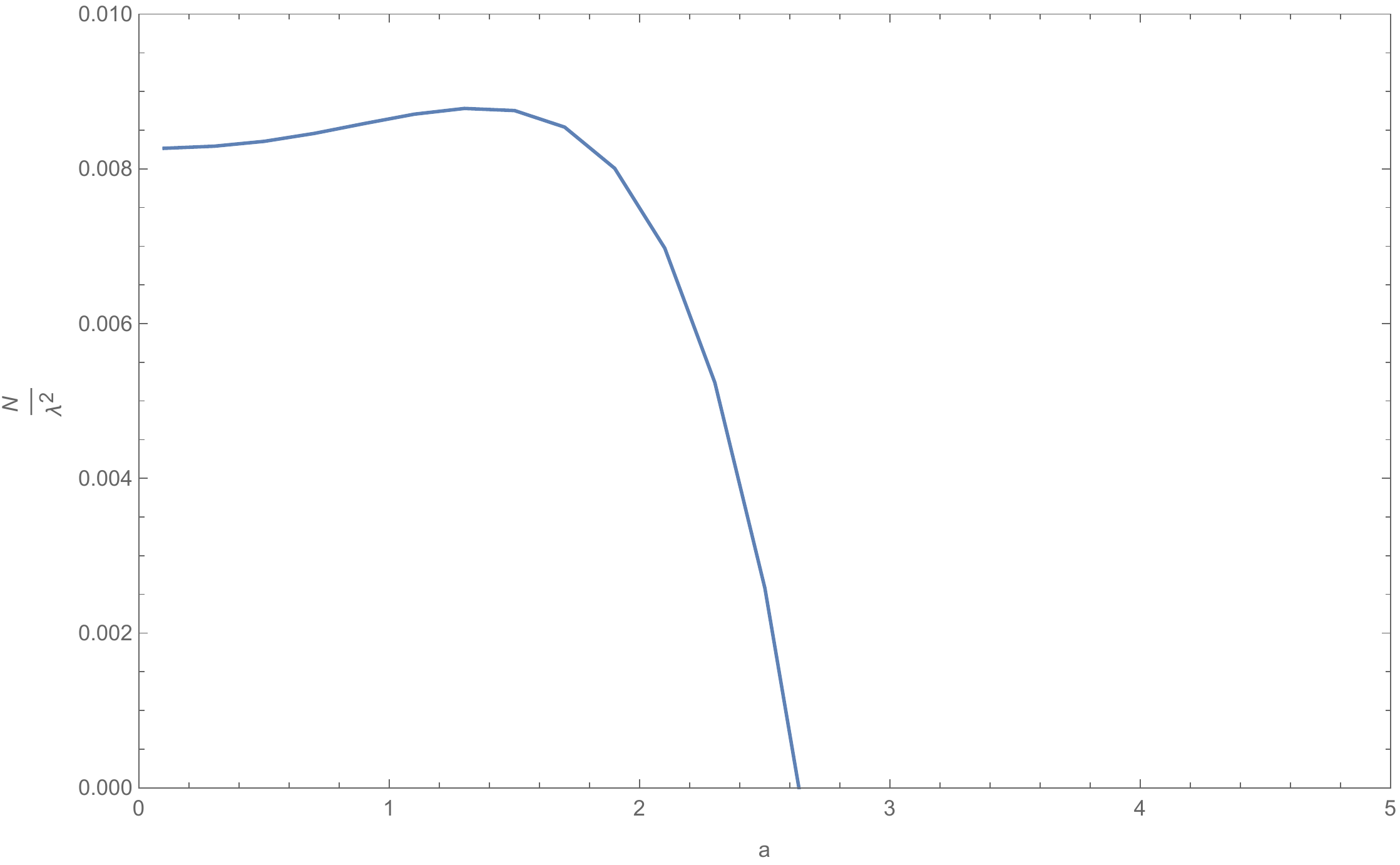}
  \caption[The logarithmic negativity of accelerating Unruh-DeWitt detectors in a Dirac field ($\sigma=1,\Omega=1,x_0=1,\Lambda=20$)]{%
The logarithmic negativity of accelerating Unruh-DeWitt detectors in a Dirac field ($\sigma=1,\Omega=1,x_0=1,\Lambda=20$)%
   }
  \label{fig:Dirac20-1}
\end{figure}

We are also interested in how the distance affects the entanglement between the detectors. Comparing figures~\ref{fig:Dirac20-0.5},~\ref{fig:Dirac20-0.7} and~\ref{fig:Dirac20-1}, we note that with some parameter values,  the entanglement decreases with the distance, which is quite intuitive.

\subsection{Mutual information}
\begin{figure}[htb!]
  \centering
  \includegraphics[width=0.78\textwidth]{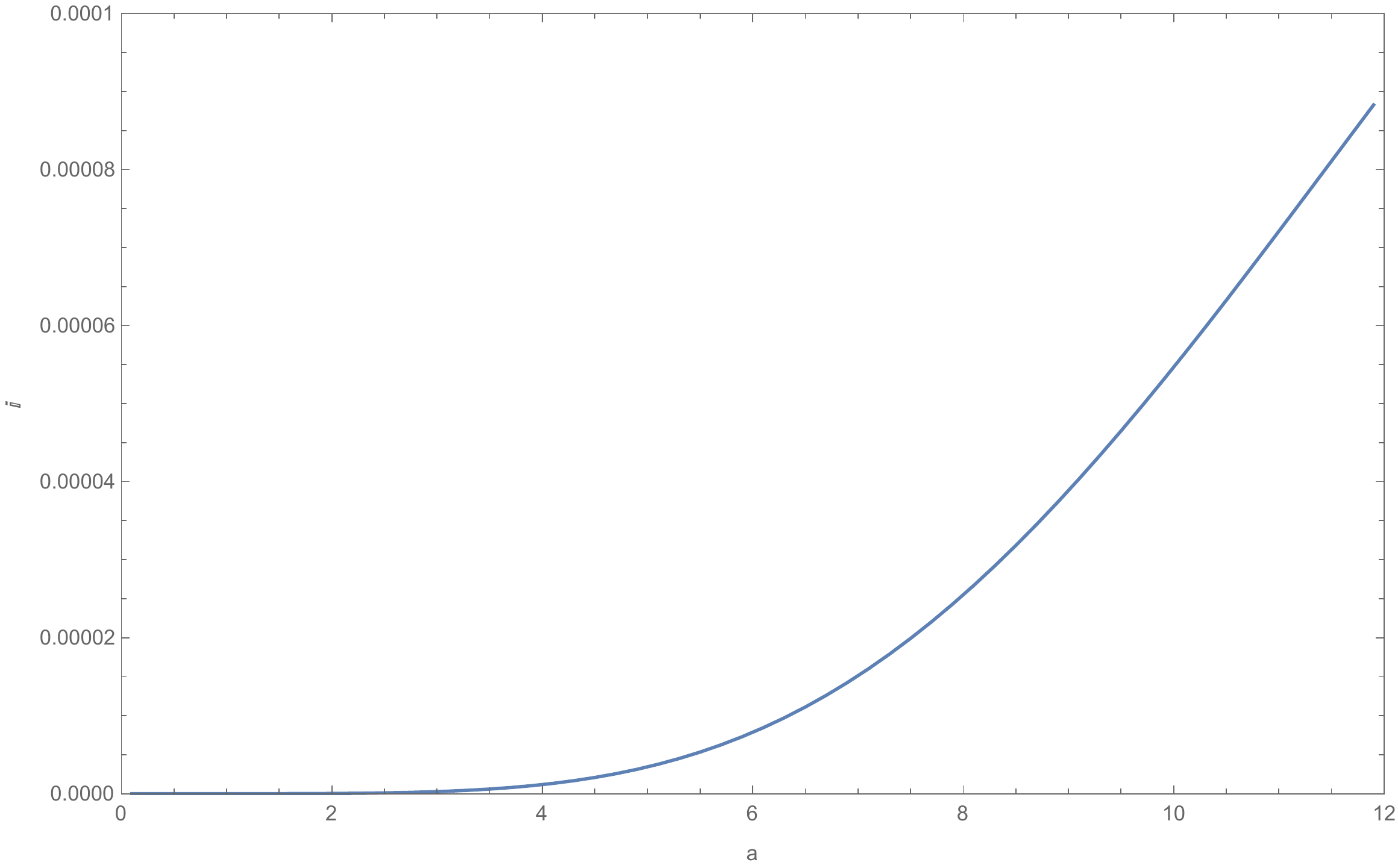}
  \caption[Mutual information between two accelerating detectors in a Dirac field ($\sigma=1,\Omega=1,x_0=0.1$)]{%
Mutual information between two accelerating detectors in a Dirac field ($\sigma=1,\Omega=1,x_0=0.1$)%
   }
  \label{fig:mutual0.1}
\end{figure}

We do not have a good description of  quantum entanglement as it involves divergent terms in the density matrix. However, mutual information, which describes the overall correlations between the detectors, is related only to $L^\Psi$ and therefore is well-defined. The expression of the mutual information for the reduced density matrix is given by~\cite{PhysRevD.96.085012}
\begin{equation}
I(\rho_{AB})=L_+\log(L_+)+L_-\log(L_-)-L_{AA}\log(L_{AA})-L_{BB}\log(L_{BB})+\mathcal{O}(\lambda^4),
\end{equation}
where
\begin{equation}
L_\pm=\frac{1}{2}\left( L_{AA}+L_{BB}\pm \sqrt{(L_{AA}-L_{BB})^2+4|L_{AB}|^2}\right).
\end{equation}
Then it can be easily seen that the mutual information is always positive due to the convexity of $x\log x$. Setting, for example, $x_0=0.1$ and $\lambda^2=0.00001$, we obtain the mutual information as in figure~\ref{fig:mutual0.1}.
It follows that the mutual information increases monotonically with the acceleration. This is because that when the two detectors coincide in spacetime, the mutual information reduces to transition probability $I(\rho)=\left(2\log 2\right) P$, which has been shown  in section \ref{sec:L} to increase  with the acceleration.

\section{Conclusion and outlook}
\label{Sec:con}
In this paper we have studied  the behavior of accelerating Unruh-DeWitt detectors in a spinor (Dirac) field in (3+1)-dimensional spacetime. For a single detector, there is a probability of transition from the ground state to the excited state after interacting with the field. We show that such probability increases with the acceleration of the detector. Therefore no antiUnruh effect  exists for point-like detectors with Gaussian switching function. 

The entanglement between the Unruh-DeWitt detectors involves divergent entries in the density matrix. The divergence comes from the properties of the Wightman function, i.e. the spinoral properties of the Dirac field. After introducing some UV cutoff, we show that logarithmic negativity of the detectors behaves nonmonotonically, as in the case of quadratic coupling with a scalar field. The mutual information, which describes the classical and quantum correlations, is independent of the divergent terms and increases with the acceleration monotonically.

Given the equivalence between the Schwarzschild metric and the Rindler metric near the horizon, our results naturally extend to the case of a black hole. In future work we wish to consider detectors which are initially entangled. As shown in ~\cite{Hummer:2015xaa}, extra terms with a time-ordered switching function  should be included. We expect our method to be valid in such case as well.

\acknowledgments
We thank Daiqin Su for useful discussions. This work was supported by National Science Foundation of China (Grant No. 12075059).

\appendix
\section{Special function and integrals}
\label{App:Special function and integrals}
$K$ is modified Bessel function of the second kind and $J$ is Bessel function of the first kind.
\begin{align}
  &\partial_{\xi}K_{i\omega/a}\left(\frac{\kappa}{a}e^{a\xi}\right)
  =-\frac{1}{2}\left[K_{i\omega/a-1}\left(\frac{\kappa}{a}e^{a\xi}\right) +K_{i\omega/a+1}\left(\frac{\kappa}{a}e^{a\xi}\right)\right]\kappa e^{a\xi}\\
  &\int_0^{2\pi}e^{\pm iu\cos\alpha}d\alpha=2\pi J_0(u)\\
  &\int_0^{2\pi}\cos\alpha e^{\pm iu\cos\alpha}d\alpha=\pm2\pi i J_1(u)\\
  &\int_0^\infty k K_{i\omega/a}(k/a)^2J_0(kx_0)dk=\frac{\pi a}{x_0\sqrt{a^2x_0^2+4}\sinh(\pi\omega/a)}\sin\frac{2\omega\text{arcsinh}(ax_0/2)}{a}\\
  &\int_0^\infty k^2 [K_{i\omega/a-1}(k/a)+K_{i\omega/a+1}(k/a)]K_{i\omega/a}(k/a)J_0(kx_0)dk=\\
  &\quad\frac{2a^2\pi}{(a^2x_0^2+4)^2\sinh(\pi\omega/a)}\left[(a^2x_0^2+4)\omega \cos\frac{2\omega\text{arcsinh}(ax_0/2)}{a}+\frac{2}{x_0}\sqrt{a^2x_0^2+4} \sin\frac{2\omega\text{arcsinh}(ax_0/2)}{a}\right]\\
  &\int_0^\infty k^2K_{i\omega/a}(k/a)^2J_1(kx_0)dk=\\
  &\frac{\frac{2\pi a}{\sinh(\pi\omega/a)}\left[-\sqrt{a^2x_0^2+4}\omega\cos\frac{2\omega\text{arccsch}(2/ax_0)}{a}+\frac{2+a^2x_0^2}{x_0}\sin\frac{2\omega\text{arccsch}(2/ax_0)}{a}\right]}{\sqrt{a^2x_0^2+4}x_0(a^2x_0^2+4)} \\
  &\int_0^\infty k K_{i\omega/a}(k/a)^2dk=\frac{\pi\omega a}{2\sinh(\pi\omega/a)}\\
  &\int_0^\infty k^2 [K_{i\omega/a-1}(k/a)+K_{i\omega/a+1}(k/a)]K_{i\omega/a}(k/a)dk= \frac{\pi\omega a^2}{\sinh(\pi\omega/a)}
\end{align}

\section{Correlation function}
\label{App:Correlation function}
Here we give a detailed proof of Eq.~\eqref{eq:2.8}.\footnote{A similar proof can be found in ~\cite{Gray:2018ifq}; however, our proof in this Appendix is intended to fit the context of the paper.}

\begin{align} 
 \notag
 V_\Psi=&\langle0_M|N\left(\bar{\Psi}_1\Psi_1\right)N\left(\bar{\Psi}_2\Psi_2\right)|0_M\rangle\\ \notag
    =&\langle0_M|\bar{\Psi}_{1a}^+\Psi_{1a}^+\bar{\Psi}_{2b}^-\Psi_{2b}^-|0_M\rangle\\ \notag
    =&\{\Psi_{1a}^+,\bar{\Psi}_{2b}^-\}\langle0_M|\bar{\Psi}_{1a}^+\Psi_{2b}^-|0_M\rangle\\ \notag
    =&iS^+_{ab}(x_1-x_2)iS^-_{ba}(x_2-x_1)\\ \notag
    =&-\text{Tr}\left[S^+(x_1-x_2)S^-(x_2-x_1)\right]\\ \notag
    =&-\text{Tr}\left[\left(i\gamma^\mu\partial_{x_1^\mu}+m\right)\Delta^+(x_1-x_2)\left(i\gamma^\nu\partial_{x_2^\nu}+m\right)\Delta^-(x_2-x_1)\right]\\ \notag
    =&\text{Tr}\left[\left(i\gamma^\mu\partial_{x_1^\mu}+m\right)\Delta^+(x_1-x_2)\left(i\gamma^\nu\partial_{x_2^\nu}+m\right)\Delta^+(x_1-x_2)\right]\\ \notag
    =&-\text{Tr}\left(\gamma^\mu\gamma^\nu\right)\partial_{x_1^\mu}\Delta^+(x_1-x_2)\partial_{x_2^\nu}\Delta^+(x_1-x_2)+4m^2\Delta^+(x_1-x_2)^2\\ \notag
    =&\text{Tr}\left(\gamma^\mu\gamma^\nu\right)\partial_{x_1^\mu}\Delta^+(x_1-x_2)\partial_{x_1^\nu}\Delta^+(x_1-x_2)+4m^2\Delta^+(x_1-x_2)^2\\ \notag
    =&-4\eta^{\mu\nu}\left(\partial_\mu\langle0_M|\hat{\Phi}_1\hat{\Phi}_2|0_M\rangle\right)\left(\partial_\nu\langle0_M|\hat{\Phi}_1\hat{\Phi}_2|0_M\rangle\right)-4m^2\langle0_M|\hat{\Phi}_1\hat{\Phi}_2|0_M\rangle^2\\ \notag
    =&-4g^{\mu'\nu'}\left(\partial_{\mu'}\langle0_M|\hat{\Phi}_1\hat{\Phi}_2|0_M\rangle\right)\left(\partial_{\nu'}\langle0_M|\hat{\Phi}_1\hat{\Phi}_2|0_M\rangle\right)-4m^2\langle0_M|\hat{\Phi}_1\hat{\Phi}_2|0_M\rangle^2\\ \notag
    =&-4e^{-2a\xi}\left(\partial_\tau\langle0_M|\hat{\Phi}_1\hat{\Phi}_2|0_M\rangle\right)^2 +4e^{-2a\xi}\left(\partial_\xi\langle0_M|\hat{\Phi}_1\hat{\Phi}_2|0_M\rangle\right)^2 \\ \notag
    &+4\left(\nabla_{\vec{x}_\bot}\langle0_M|\hat{\Phi}_1\hat{\Phi}_2|0_M\rangle\right)^2 -4m^2\langle0_M|\hat{\Phi}_1\hat{\Phi}_2|0_M\rangle^2\\
    \equiv& -4e^{-2a\xi}V_\Psi^\tau+4e^{-2a\xi}V_\Psi^\xi+4V_\Psi^\bot-4m^2V_\Psi^m,
\end{align}
where $S_{ab}$ is the Dirac propagator~\cite{Mandl:1985bg} and the Wightman function for scalar fields is related to $\Delta^+(x-y)$ as
\begin{equation}
i\Delta^+(x-y)=W(x-y).
\end{equation}

\section{Calculation of $L$}
\label{App: Calculation of $L$}
In this section we display the results of the calculation.
\begin{align}
I^{\tau1}_{\omega_1\vec{k}_{1\bot},\omega_2\vec{k}_{2\bot}}
  =&\omega_1\omega_2\eta^{1A'*}_{\omega_1\vec{k}_{1\bot},\omega_2\vec{k}_{2\bot}}\eta^{1B}_{\omega_1\vec{k}_{1\bot},\omega_2\vec{k}_{2\bot}},\\
I^{\tau2}_{\omega_1\vec{k}_{1\bot},\omega_2\vec{k}_{2\bot}}
  =&\omega_1\omega_2\eta^{2A'*}_{\omega_1\vec{k}_{1\bot},\omega_2\vec{k}_{2\bot}}\eta^{2B}_{\omega_1\vec{k}_{1\bot},\omega_2\vec{k}_{2\bot}},\\
I^{\tau12}_{\omega_1\vec{k}_{1\bot},\omega_2\vec{k}_{2\bot}}
  =&-\omega_1\omega_2\eta^{12A'*}_{\omega_1\vec{k}_{1\bot},\omega_2\vec{k}_{2\bot}}\eta^{12B}_{\omega_1\vec{k}_{1\bot},\omega_2\vec{k}_{2\bot}}.
\end{align}

\begin{align} \notag
I^{\vec{x}_\bot 0}_{\omega_1\vec{k}_{1\bot},\omega_2\vec{k}_{2\bot}}
    =&\int d\tau_1\int d\tau_2  \chi_{A1} \chi_{B2} e^{-i\Omega_A\tau_1+i\Omega_B\tau_2}\int_{\Sigma_{A1}}\widetilde{d}^3\vec{x}_1f_{A1}^*\int_{\Sigma_{B2}}\widetilde{d}^3\vec{x}_2f_{B2}\\\notag
    &\times\left(\nabla_{\vec{x}_{1\bot}}v^R_{1\omega_1\vec{k}_{1\bot}}\right)\left(\nabla_{\vec{x}_{1\bot}}v^R_{1\omega_2\vec{k}_{2\bot}}\right)v^{R*}_{2\omega_1\vec{k}_{1\bot}}v^{R*}_{2\omega_2\vec{k}_{2\bot}}\\\notag
    =&-\vec{k}_{1\bot}\cdot\vec{k}_{2\bot}\int d\tau_1\int d\tau_2  \chi_{A1} \chi_{B2} e^{-i\Omega_A\tau_1+i\Omega_B\tau_2}\int_{\Sigma_{A1}}\widetilde{d}^3\vec{x}_1f_{A1}^*\int_{\Sigma_{B2}}\widetilde{d}^3\vec{x}_2f_{B2}\\\notag
    &\times v^R_{1\omega_1\vec{k}_{1\bot}}v^R_{1\omega_2\vec{k}_{2\bot}}v^{R*}_{2\omega_1\vec{k}_{1\bot}}v^{R*}_{2\omega_2\vec{k}_{2\bot}}\\
    =&-\vec{k}_{1\bot}\cdot\vec{k}_{2\bot}\eta^{0A*}_{\omega_1\vec{k}_{1\bot},\omega_2\vec{k}_{2\bot}}\eta^{0B}_{\omega_1\vec{k}_{1\bot},\omega_2\vec{k}_{2\bot}}.
\end{align}

\begin{align}
I^{\vec{x}_\bot 1}_{\omega_1\vec{k}_{1\bot},\omega_2\vec{k}_{2\bot}}
    =&\vec{k}_{1\bot}\cdot\vec{k}_{2\bot}\eta^{1A*}_{\omega_1\vec{k}_{1\bot},\omega_2\vec{k}_{2\bot}}\eta^{1B}_{\omega_1\vec{k}_{1\bot},\omega_2\vec{k}_{2\bot}},\\
I^{\vec{x}_\bot 2}_{\omega_1\vec{k}_{1\bot},\omega_2\vec{k}_{2\bot}}
    =&\vec{k}_{1\bot}\cdot\vec{k}_{2\bot}\eta^{2A*}_{\omega_1\vec{k}_{1\bot},\omega_2\vec{k}_{2\bot}}\eta^{2B}_{\omega_1\vec{k}_{1\bot},\omega_2\vec{k}_{2\bot}},\\
I^{\vec{x}_\bot 12}_{\omega_1\vec{k}_{1\bot},\omega_2\vec{k}_{2\bot}}
    =&-\vec{k}_{1\bot}\cdot\vec{k}_{2\bot}\eta^{12A*}_{\omega_1\vec{k}_{1\bot},\omega_2\vec{k}_{2\bot}}\eta^{12B}_{\omega_1\vec{k}_{1\bot},\omega_2\vec{k}_{2\bot}}.
\end{align}

\begin{align}\notag
I^{\xi1}_{\omega_1\vec{k}_{1\bot},\omega_2\vec{k}_{2\bot}}
  =& \int d\tau_1\int d\tau_2  \chi_{A1} \chi_{B2} e^{-i\Omega_A\tau_1+i\Omega_B\tau_2}\int_{\Sigma_{A1}}e^{-2a\xi_1}\widetilde{d}^3\vec{x}_1f_{A1}^*\int_{\Sigma_{B2}}\widetilde{d}^3\vec{x}_2f_{B2}\\\notag
  &\times \left(\partial_{\xi_1}v^{R*}_{1\omega_1\vec{k}_{1\bot}}\right)\left(\partial_{\xi_1}v^R_{1\omega_2\vec{k}_{2\bot}}\right)v^R_{2\omega_1\vec{k}_{1\bot}}v^{R*}_{2\omega_2\vec{k}_{2\bot}}\\\notag
  =&\frac{\sinh(\pi\omega_1/a)\sinh(\pi\omega_2/a)}{4(4\pi^4a)^2/(2\pi)}G_B(\Omega_B-\omega_1+\omega_2)G_A(-\Omega_A+\omega_1-\omega_2) \\\notag
  &\times\int_{\Sigma_{A1}}\widetilde{d}^3\vec{x}_1f_{A1}^*\int_{\Sigma_{B2}}\widetilde{d}^3\vec{x}_2f_{B2}
  e^{i(-\vec{k}_{1\bot}+\vec{k}_{2\bot})\cdot(\vec{x}_{1\bot}-\vec{x}_{2\bot})}\\\notag
  &\hspace{-3cm}\times\kappa_1\kappa_2\left[K_{i\omega_1/a-1}\left(\frac{\kappa_1}{a}e^{a\xi_1}\right) +K_{i\omega_1/a+1}\left(\frac{\kappa_1}{a}e^{a\xi_1}\right)\right] \left[K_{i\omega_2/a-1}\left(\frac{\kappa_2}{a}e^{a\xi_1}\right) +K_{i\omega_2/a+1}\left(\frac{\kappa_2}{a}e^{a\xi_1}\right)\right] \\
  &\times K_{i\omega_1/a}\left(\frac{\kappa_1}{a}e^{a\xi_2}\right) K_{i\omega_2/a}\left(\frac{\kappa_2}{a}e^{a\xi_2}\right),\\\notag
I^{\xi2}_{\omega_1\vec{k}_{1\bot},\omega_2\vec{k}_{2\bot}}
  =& \int d\tau_1\int d\tau_2  \chi_{A1} \chi_{B2} e^{-i\Omega_A\tau_1+i\Omega_B\tau_2}\int_{\Sigma_{A1}}e^{-2a\xi_1}\widetilde{d}^3\vec{x}_1f_{A1}^*\int_{\Sigma_{B2}}\widetilde{d}^3\vec{x}_2f_{B2}\\\notag
  &\times \left(\partial_{\xi_1}v^R_{1\omega_1\vec{k}_{1\bot}}\right)\left(\partial_{\xi_1}v^{R*}_{1\omega_2\vec{k}_{2\bot}}\right)v^{R*}_{2\omega_1\vec{k}_{1\bot}}v^R_{2\omega_2\vec{k}_{2\bot}}\\\notag
  =&\frac{\sinh(\pi\omega_1/a)\sinh(\pi\omega_2/a)}{4(4\pi^4a)^2/(2\pi)}G_B(\Omega_B+\omega_1-\omega_2)G_A(-\Omega_A-\omega_1+\omega_2) \\\notag
  &\times\int_{\Sigma_{A1}}\widetilde{d}^3\vec{x}_1f_{A1}^*\int_{\Sigma_{B2}}\widetilde{d}^3\vec{x}_2f_{B2}
  e^{i(\vec{k}_{1\bot}-\vec{k}_{2\bot})\cdot(\vec{x}_{1\bot}-\vec{x}_{2\bot})}\\\notag
  &\hspace{-3cm}\times\kappa_1\kappa_2\left[K_{i\omega_1/a-1}\left(\frac{\kappa_1}{a}e^{a\xi_1}\right) +K_{i\omega_1/a+1}\left(\frac{\kappa_1}{a}e^{a\xi_1}\right)\right] \left[K_{i\omega_2/a-1}\left(\frac{\kappa_2}{a}e^{a\xi_1}\right) +K_{i\omega_2/a+1}\left(\frac{\kappa_2}{a}e^{a\xi_1}\right)\right] \\
  &\times K_{i\omega_1/a}\left(\frac{\kappa_1}{a}e^{a\xi_2}\right) K_{i\omega_2/a}\left(\frac{\kappa_2}{a}e^{a\xi_2}\right),\\\notag
I^{\xi12}_{\omega_1\vec{k}_{1\bot},\omega_2\vec{k}_{2\bot}}
  =& \int d\tau_1\int d\tau_2  \chi_{A1} \chi_{B2} e^{-i\Omega_A\tau_1+i\Omega_B\tau_2}\int_{\Sigma_{A1}}e^{-2a\xi_1}\widetilde{d}^3\vec{x}_1f_{A1}^*\int_{\Sigma_{B2}}\widetilde{d}^3\vec{x}_2f_{B2}\\\notag
  &\times \left(\partial_{\xi_1}v^{R*}_{1\omega_1\vec{k}_{1\bot}}\right)\left(\partial_{\xi_1}v^{R*}_{1\omega_2\vec{k}_{2\bot}}\right)v^R_{2\omega_1\vec{k}_{1\bot}}v^R_{2\omega_2\vec{k}_{2\bot}}\\\notag
  =&\frac{\sinh(\pi\omega_1/a)\sinh(\pi\omega_2/a)}{4(4\pi^4a)^2/(2\pi)}G_B(\Omega_B-\omega_1-\omega_2)G_A(-\Omega_A+\omega_1+\omega_2) \\\notag
  &\times\int_{\Sigma_{A1}}\widetilde{d}^3\vec{x}_1f_{A1}^*\int_{\Sigma_{B2}}\widetilde{d}^3\vec{x}_2f_{B2}
  e^{i(-\vec{k}_{1\bot}-\vec{k}_{2\bot})\cdot(\vec{x}_{1\bot}-\vec{x}_{2\bot})}\\\notag
  &\hspace{-3cm}\times\kappa_1\kappa_2\left[K_{i\omega_1/a-1}\left(\frac{\kappa_1}{a}e^{a\xi_1}\right) +K_{i\omega_1/a+1}\left(\frac{\kappa_1}{a}e^{a\xi_1}\right)\right] \left[K_{i\omega_2/a-1}\left(\frac{\kappa_2}{a}e^{a\xi_1}\right) +K_{i\omega_2/a+1}\left(\frac{\kappa_2}{a}e^{a\xi_1}\right)\right] \\
  &\times K_{i\omega_1/a}\left(\frac{\kappa_1}{a}e^{a\xi_2}\right) K_{i\omega_2/a}\left(\frac{\kappa_2}{a}e^{a\xi_2}\right).
\end{align}

Assuming point-like detectors and Gaussian switching functions, we have the following results.

\begin{align}\notag
I^{\tau0}_{\omega_1\vec{k}_{1\bot},\omega_2\vec{k}_{2\bot}} =&-\omega_1\omega_2\frac{\sinh(\pi\omega_1/a)\sinh(\pi\omega_2/a)}{(4\pi^4a)^2/(2\pi)}
    K_{i\omega_1/a}\left(\frac{\kappa_1}{a}\right)^2 K_{i\omega_2/a}\left(\frac{\kappa_2}{a}\right)^2\\\notag
    &\times\exp\Bigg[-\frac{\sigma_A^2}{2}(\Omega_A+\omega_1+\omega_2)^2-\frac{\sigma_B^2}{2}(\Omega_B+\omega_1+\omega_2)^2\\
    &+i(\Omega_B+\omega_1+\omega_2)\tau_0-i(k_{1x}+k_{2x})x_0\Bigg],\\\notag
I^{\tau1}_{\omega_1\vec{k}_{1\bot},\omega_2\vec{k}_{2\bot}} =&\omega_1\omega_2\frac{\sinh(\pi\omega_1/a)\sinh(\pi\omega_2/a)}{(4\pi^4a)^2/(2\pi)}
    K_{i\omega_1/a}\left(\frac{\kappa_1}{a}\right)^2 K_{i\omega_2/a}\left(\frac{\kappa_2}{a}\right)^2\\\notag
    &\times\exp\Bigg[-\frac{\sigma_A^2}{2}(\Omega_A-\omega_1+\omega_2)^2-\frac{\sigma_B^2}{2}(\Omega_B-\omega_1+\omega_2)^2\\
    &+i(\Omega_B-\omega_1+\omega_2)\tau_0-i(-k_{1x}+k_{2x})x_0\Bigg],\\\notag
I^{\tau2}_{\omega_1\vec{k}_{1\bot},\omega_2\vec{k}_{2\bot}} =&\omega_1\omega_2\frac{\sinh(\pi\omega_1/a)\sinh(\pi\omega_2/a)}{(4\pi^4a)^2/(2\pi)}
    K_{i\omega_1/a}\left(\frac{\kappa_1}{a}\right)^2 K_{i\omega_2/a}\left(\frac{\kappa_2}{a}\right)^2\\\notag
    &\times\exp\Bigg[-\frac{\sigma_A^2}{2}(\Omega_A+\omega_1-\omega_2)^2-\frac{\sigma_B^2}{2}(\Omega_B+\omega_1-\omega_2)^2\\
    &+i(\Omega_B+\omega_1-\omega_2)\tau_0-i(k_{1x}-k_{2x})x_0\Bigg],\\\notag
I^{\tau12}_{\omega_1\vec{k}_{1\bot},\omega_2\vec{k}_{2\bot}} =&-\omega_1\omega_2\frac{\sinh(\pi\omega_1/a)\sinh(\pi\omega_2/a)}{(4\pi^4a)^2/(2\pi)}
    K_{i\omega_1/a}\left(\frac{\kappa_1}{a}\right)^2 K_{i\omega_2/a}\left(\frac{\kappa_2}{a}\right)^2\\\notag
    &\times\exp\Bigg[-\frac{\sigma_A^2}{2}(\Omega_A-\omega_1-\omega_2)^2-\frac{\sigma_B^2}{2}(\Omega_B-\omega_1-\omega_2)^2\\
    &+i(\Omega_B-\omega_1-\omega_2)\tau_0-i(-k_{1x}-k_{2x})x_0\Bigg].
\end{align}

\begin{align}\notag
I^{x_\bot 0}_{\omega_1\vec{k}_{1\bot},\omega_2\vec{k}_{2\bot}} =&-\vec{k}_{1\bot}\cdot\vec{k}_{2\bot}\frac{\sinh(\pi\omega_1/a)\sinh(\pi\omega_2/a)}{(4\pi^4a)^2/(2\pi)}
    K_{i\omega_1/a}\left(\frac{\kappa_1}{a}\right)^2 K_{i\omega_2/a}\left(\frac{\kappa_2}{a}\right)^2\\\notag
    &\times\exp\Bigg[-\frac{\sigma_A^2}{2}(\Omega_A+\omega_1+\omega_2)^2-\frac{\sigma_B^2}{2}(\Omega_B+\omega_1+\omega_2)^2\\
    &+i(\Omega_B+\omega_1+\omega_2)\tau_0-i(k_{1x}+k_{2x})x_0\Bigg],\\\notag
I^{x_\bot 1}_{\omega_1\vec{k}_{1\bot},\omega_2\vec{k}_{2\bot}} =&\vec{k}_{1\bot}\cdot\vec{k}_{2\bot}\frac{\sinh(\pi\omega_1/a)\sinh(\pi\omega_2/a)}{(4\pi^4a)^2/(2\pi)}
    K_{i\omega_1/a}\left(\frac{\kappa_1}{a}\right)^2 K_{i\omega_2/a}\left(\frac{\kappa_2}{a}\right)^2\\\notag
    &\times\exp\Bigg[-\frac{\sigma_A^2}{2}(\Omega_A-\omega_1+\omega_2)^2-\frac{\sigma_B^2}{2}(\Omega_B-\omega_1+\omega_2)^2\\
    &+i(\Omega_B-\omega_1+\omega_2)\tau_0-i(-k_{1x}+k_{2x})x_0\Bigg],\\\notag
I^{x_\bot 2}_{\omega_1\vec{k}_{1\bot},\omega_2\vec{k}_{2\bot}} =&\vec{k}_{1\bot}\cdot\vec{k}_{2\bot}\frac{\sinh(\pi\omega_1/a)\sinh(\pi\omega_2/a)}{(4\pi^4a)^2/(2\pi)}
    K_{i\omega_1/a}\left(\frac{\kappa_1}{a}\right)^2 K_{i\omega_2/a}\left(\frac{\kappa_2}{a}\right)^2\\\notag
    &\times\exp\Bigg[-\frac{\sigma_A^2}{2}(\Omega_A+\omega_1-\omega_2)^2-\frac{\sigma_B^2}{2}(\Omega_B+\omega_1-\omega_2)^2\\
    &+i(\Omega_B+\omega_1-\omega_2)\tau_0-i(k_{1x}-k_{2x})x_0\Bigg],\\\notag
I^{x_\bot 12}_{\omega_1\vec{k}_{1\bot},\omega_2\vec{k}_{2\bot}} =&-\vec{k}_{1\bot}\cdot\vec{k}_{2\bot}\frac{\sinh(\pi\omega_1/a)\sinh(\pi\omega_2/a)}{(4\pi^4a)^2/(2\pi)}
    K_{i\omega_1/a}\left(\frac{\kappa_1}{a}\right)^2 K_{i\omega_2/a}\left(\frac{\kappa_2}{a}\right)^2\\\notag
    &\times\exp\Bigg[-\frac{\sigma_A^2}{2}(\Omega_A-\omega_1-\omega_2)^2-\frac{\sigma_B^2}{2}(\Omega_B-\omega_1-\omega_2)^2\\
    &+i(\Omega_B-\omega_1-\omega_2)\tau_0-i(-k_{1x}-k_{2x})x_0\Bigg].
\end{align}

\begin{align}\notag
I^{\xi0}_{\omega_1\vec{k}_{1\bot},\omega_2\vec{k}_{2\bot}}
    =&\kappa_1\kappa_2\frac{\sinh(\pi\omega_1/a)\sinh(\pi\omega_2/a)}{4(4\pi^4a)^2/(2\pi)}\\\notag
    &\times\exp\Bigg[-\frac{\sigma_A^2}{2}(\Omega_A+\omega_1+\omega_2)^2-\frac{\sigma_B^2}{2}(\Omega_B+\omega_1+\omega_2)^2\\\notag
    &+i(\Omega_B+\omega_1+\omega_2)\tau_0-i(k_{1x}+k_{2x})x_0\Bigg]\\\notag
    &\times\left[K_{i\omega_1/a-1}\left(\frac{\kappa_1}{a}\right) +K_{i\omega_1/a+1}\left(\frac{\kappa_1}{a}\right)\right] \left[K_{i\omega_2/a-1}\left(\frac{\kappa_2}{a}\right) +K_{i\omega_2/a+1}\left(\frac{\kappa_2}{a}\right)\right]\\
    &\times  K_{i\omega_1/a}\left(\frac{\kappa_1}{a}\right) K_{i\omega_2/a}\left(\frac{\kappa_2}{a}\right),\\\notag
I^{\xi1}_{\omega_1\vec{k}_{1\bot},\omega_2\vec{k}_{2\bot}}
    =&\kappa_1\kappa_2\frac{\sinh(\pi\omega_1/a)\sinh(\pi\omega_2/a)}{4(4\pi^4a)^2/(2\pi)}\\\notag
    &\times\exp\Bigg[-\frac{\sigma_A^2}{2}(\Omega_A-\omega_1+\omega_2)^2-\frac{\sigma_B^2}{2}(\Omega_B-\omega_1+\omega_2)^2\\\notag
    &+i(\Omega_B-\omega_1+\omega_2)\tau_0-i(-k_{1x}+k_{2x})x_0\Bigg]\\\notag
    &\times\left[K_{i\omega_1/a-1}\left(\frac{\kappa_1}{a}\right) +K_{i\omega_1/a+1}\left(\frac{\kappa_1}{a}\right)\right] \left[K_{i\omega_2/a-1}\left(\frac{\kappa_2}{a}\right) +K_{i\omega_2/a+1}\left(\frac{\kappa_2}{a}\right)\right]\\
    &\times K_{i\omega_1/a}\left(\frac{\kappa_1}{a}\right) K_{i\omega_2/a}\left(\frac{\kappa_2}{a}\right),\\\notag
I^{\xi2}_{\omega_1\vec{k}_{1\bot},\omega_2\vec{k}_{2\bot}}
    =&\kappa_1\kappa_2\frac{\sinh(\pi\omega_1/a)\sinh(\pi\omega_2/a)}{4(4\pi^4a)^2/(2\pi)}\\\notag
    &\times\exp\Bigg[-\frac{\sigma_A^2}{2}(\Omega_A+\omega_1-\omega_2)^2-\frac{\sigma_B^2}{2}(\Omega_B+\omega_1-\omega_2)^2\\\notag
    &+i(\Omega_B+\omega_1-\omega_2)\tau_0-i(k_{1x}-k_{2x})x_0\Bigg]\\\notag
    &\times\left[K_{i\omega_1/a-1}\left(\frac{\kappa_1}{a}\right) +K_{i\omega_1/a+1}\left(\frac{\kappa_1}{a}\right)\right] \left[K_{i\omega_2/a-1}\left(\frac{\kappa_2}{a}\right) +K_{i\omega_2/a+1}\left(\frac{\kappa_2}{a}\right)\right] \\
    &\times K_{i\omega_1/a}\left(\frac{\kappa_1}{a}\right) K_{i\omega_2/a}\left(\frac{\kappa_2}{a}\right),\\\notag
I^{\xi12}_{\omega_1\vec{k}_{1\bot},\omega_2\vec{k}_{2\bot}}
    =&\kappa_1\kappa_2\frac{\sinh(\pi\omega_1/a)\sinh(\pi\omega_2/a)}{4(4\pi^4a)^2/(2\pi)}\\\notag
    &\times\exp\Bigg[-\frac{\sigma_A^2}{2}(\Omega_A-\omega_1-\omega_2)^2-\frac{\sigma_B^2}{2}(\Omega_B-\omega_1-\omega_2)^2\\\notag
    &+i(\Omega_B-\omega_1-\omega_2)\tau_0-i(-k_{1x}-k_{2x})x_0\Bigg]\\\notag
    &\times\left[K_{i\omega_1/a-1}\left(\frac{\kappa_1}{a}\right) +K_{i\omega_1/a+1}\left(\frac{\kappa_1}{a}\right)\right] \left[K_{i\omega_2/a-1}\left(\frac{\kappa_2}{a}\right) +K_{i\omega_2/a+1}\left(\frac{\kappa_2}{a}\right)\right] \\
    &\times K_{i\omega_1/a}\left(\frac{\kappa_1}{a}\right) K_{i\omega_2/a}\left(\frac{\kappa_2}{a}\right).
\end{align}

Now we use polar coordinates to integrate $k_\bot$.

\begin{align}\notag
 L^{\Psi\vec{x}_\bot}_{AB}=&\frac{2\pi}{(4\pi^4a)^2}\int_0^\infty d\omega_1\int_0^\infty d\omega_2\iint d^2\vec{k}_{1\bot}
    \iint d^2\vec{k}_{2\bot} \vec{k}_{1\bot}\cdot\vec{k}_{2\bot} \\\notag
    &\times \frac{\sinh(\pi\omega_1/a)}{1-e^{-2\pi\omega_1/a}}\frac{\sinh(\pi\omega_2/a)}{1-e^{-2\pi\omega_2/a}} K_{i\omega_1/a}\left(\frac{\kappa_1}{a}\right)^2K_{i\omega_2/a}\left(\frac{\kappa_2}{a}\right)^2\\\notag
    &\times\left\{-\exp\left[-\frac{\sigma_A^2}{2}(\omega_1+\omega_2+\Omega_A)^2-\frac{\sigma_B^2}{2}(\omega_1+\omega_2+\Omega_B)^2-i(k_{1x}+k_{2x})x_0\right]\right.\\\notag
    &+\exp\left[-\frac{\sigma_A^2}{2}(\Omega_A+\omega_1-\omega_2)^2-\frac{\sigma_B^2}{2}(\Omega_B+\omega_1-\omega_2)^2-i(k_{1x}-k_{2x})x_0-2\pi\omega_2/a\right]\\\notag
    &+\exp\left[-\frac{\sigma_A^2}{2}(\Omega_A-\omega_1+\omega_2)^2-\frac{\sigma_B^2}{2}(\Omega_B-\omega_1+\omega_2)^2+i(k_{1x}-k_{2x})x_0-2\pi\omega_1/a\right]\\\notag
    &\hspace{-1cm}\left.-\exp\left[-\frac{\sigma_A^2}{2}(\Omega_A-\omega_1-\omega_2)^2-\frac{\sigma_B^2}{2}(\Omega_B-\omega_1-\omega_2)^2+i(k_{1x}+k_{2x})x_0-2\pi(\omega_1+\omega_2)/a\right]\right\}\\\notag
    =&\frac{(2\pi)^3}{4(4\pi^4a)^2}\int_0^\infty d\omega_1\int_0^\infty d\omega_2\int k_{1\bot}^2dk_{1\bot}\int k_{2\bot}^2dk_{2\bot}\\\notag
    &\times K_{i\omega_1/a}\left(\frac{\sqrt{k_{1\bot}^2+m^2}}{a}\right)^2 K_{i\omega_2/a}\left(\frac{\sqrt{k_{2\bot}^2+m^2}}{a}\right)^2J_1(k_{1\bot}x_0)J_1(k_{2\bot}x_0)\\\notag
    &\hspace{-1cm}\times\left\{\exp\left[-\frac{\sigma_A^2}{2}(\omega_1+\omega_2+\Omega_A)^2-\frac{\sigma_B^2}{2}(\omega_1+\omega_2+\Omega_B)^2+i(\omega_1+\omega_2+\Omega_B)\tau_0+\pi(\omega_1+\omega_2)/a\right]\right.\\\notag
    &\hspace{-1cm}+\exp\left[-\frac{\sigma_A^2}{2}(\Omega_A+\omega_1-\omega_2)^2-\frac{\sigma_B^2}{2}(\Omega_B+\omega_1-\omega_2)^2+i(\Omega_B+\omega_1-\omega_2)\tau_0+\pi(\omega_1-\omega_2)/a\right]\\\notag
    &\hspace{-1cm}+\exp\left[-\frac{\sigma_A^2}{2}(\Omega_A-\omega_1+\omega_2)^2-\frac{\sigma_B^2}{2}(\Omega_B-\omega_1+\omega_2)^2+i(\Omega_B-\omega_1+\omega_2)\tau_0-\pi(\omega_1-\omega_2)/a\right]\\\notag
    &\hspace{-1cm}\left.+\exp\left[-\frac{\sigma_A^2}{2}(\Omega_A-\omega_1-\omega_2)^2-\frac{\sigma_B^2}{2}(\Omega_B-\omega_1-\omega_2)^2+i(\Omega_B-\omega_1-\omega_2)\tau_0-\pi(\omega_1+\omega_2)/a\right]\right\}\\\notag
    =&\frac{(2\pi)^3}{4(4\pi^4a)^2}\int_0^\infty d\omega_1\int_0^\infty d\omega_2F(\omega_1,x_0,a)F(\omega_2,x_0,a)\\\notag
    &\hspace{-1cm}\times\left\{\exp\left[-\frac{\sigma_A^2}{2}(\omega_1+\omega_2+\Omega_A)^2-\frac{\sigma_B^2}{2}(\omega_1+\omega_2+\Omega_B)^2+i(\omega_1+\omega_2+\Omega_B)\tau_0+\pi(\omega_1+\omega_2)/a\right]\right.\\\notag
    &\hspace{-1cm}+\exp\left[-\frac{\sigma_A^2}{2}(\Omega_A+\omega_1-\omega_2)^2-\frac{\sigma_B^2}{2}(\Omega_B+\omega_1-\omega_2)^2+i(\Omega_B+\omega_1-\omega_2)\tau_0+\pi(\omega_1-\omega_2)/a\right]\\\notag
    &\hspace{-1cm}+\exp\left[-\frac{\sigma_A^2}{2}(\Omega_A-\omega_1+\omega_2)^2-\frac{\sigma_B^2}{2}(\Omega_B-\omega_1+\omega_2)^2+i(\Omega_B-\omega_1+\omega_2)\tau_0-\pi(\omega_1-\omega_2)/a\right]\\
    &\hspace{-1cm}\left.+\exp\left[-\frac{\sigma_A^2}{2}(\Omega_A-\omega_1-\omega_2)^2-\frac{\sigma_B^2}{2}(\Omega_B-\omega_1-\omega_2)^2+i(\Omega_B-\omega_1-\omega_2)\tau_0-\pi(\omega_1+\omega_2)/a\right]\right\}.
\end{align}

\begin{align}\notag
L_{AB}^{\Psi\xi}=&\frac{2\pi}{4(4\pi^4a)^2}\int_0^\infty d\omega_1\int_0^\infty d\omega_2\iint d^2\vec{k}_{1\bot}
    \iint d^2\vec{k}_{2\bot} \kappa_1\kappa_2\frac{\sinh(\pi\omega_1/a)}{1-e^{-2\pi\omega_1/a}}\frac{\sinh(\pi\omega_2/a)}{1-e^{-2\pi\omega_2/a}}\\\notag
     &\hspace{-1cm}\times\left[K_{i\omega_1/a-1}\left(\frac{\kappa_1}{a}\right) +K_{i\omega_1/a+1}\left(\frac{\kappa_1}{a}\right)\right] \left[K_{i\omega_2/a-1}\left(\frac{\kappa_2}{a}\right) +K_{i\omega_2/a+1}\left(\frac{\kappa_2}{a}\right)\right] K_{i\omega_1/a}\left(\frac{\kappa_1}{a}\right) K_{i\omega_2/a}\left(\frac{\kappa_2}{a}\right)\\\notag
    &\times\left\{\exp\left[-\frac{\sigma_A^2}{2}(\omega_1+\omega_2+\Omega_A)^2-\frac{\sigma_B^2}{2}(\omega_1+\omega_2+\Omega_B)^2-i(k_{1x}+k_{2x})x_0\right]\right.\\\notag
    &+\exp\left[-\frac{\sigma_A^2}{2}(\Omega_A+\omega_1-\omega_2)^2-\frac{\sigma_B^2}{2}(\Omega_B+\omega_1-\omega_2)^2-i(k_{1x}-k_{2x})x_0-2\pi\omega_2/a\right]\\\notag
    &+\exp\left[-\frac{\sigma_A^2}{2}(\Omega_A-\omega_1+\omega_2)^2-\frac{\sigma_B^2}{2}(\Omega_B-\omega_1+\omega_2)^2+i(k_{1x}-k_{2x})x_0-2\pi\omega_1/a\right]\\\notag
    &\hspace{-1cm}\left.+\exp\left[-\frac{\sigma_A^2}{2}(\Omega_A-\omega_1-\omega_2)^2-\frac{\sigma_B^2}{2}(\Omega_B-\omega_1-\omega_2)^2+i(k_{1x}+k_{2x})x_0-2\pi(\omega_1+\omega_2)/a\right]\right\}\\\notag
    =&\frac{(2\pi)^3}{4^2(4\pi^4a)^2}\int_0^\infty d\omega_1\int_0^\infty d\omega_2 \int k_{1\bot}dk_{1\bot}\int k_{2\bot}dk_{2\bot}\sqrt{k_{1\bot}^2}\sqrt{k_{2\bot}^2}J_0(k_{1\bot}x_0)J_0(k_{2\bot}x_0)\\\notag
     &\hspace{-1cm}\times\left[K_{i\omega_1/a-1}\left(\frac{\kappa_1}{a}\right) +K_{i\omega_1/a+1}\left(\frac{\kappa_1}{a}\right)\right] \left[K_{i\omega_2/a-1}\left(\frac{\kappa_2}{a}\right) +K_{i\omega_2/a+1}\left(\frac{\kappa_2}{a}\right)\right] K_{i\omega_1/a}\left(\frac{\kappa_1}{a}\right) K_{i\omega_2/a}\left(\frac{\kappa_2}{a}\right)\\\notag
    &\hspace{-1cm}\times\left\{\exp\left[-\frac{\sigma_A^2}{2}(\omega_1+\omega_2+\Omega_A)^2-\frac{\sigma_B^2}{2}(\omega_1+\omega_2+\Omega_B)^2+i(\omega_1+\omega_2+\Omega_B)\tau_0+\pi(\omega_1+\omega_2)/a\right]\right.\\\notag
    &\hspace{-1cm}+\exp\left[-\frac{\sigma_A^2}{2}(\Omega_A+\omega_1-\omega_2)^2-\frac{\sigma_B^2}{2}(\Omega_B+\omega_1-\omega_2)^2+i(\Omega_B+\omega_1-\omega_2)\tau_0+\pi(\omega_1-\omega_2)/a\right]\\\notag
    &\hspace{-1cm}+\exp\left[-\frac{\sigma_A^2}{2}(\Omega_A-\omega_1+\omega_2)^2-\frac{\sigma_B^2}{2}(\Omega_B-\omega_1+\omega_2)^2+i(\Omega_B-\omega_1+\omega_2)\tau_0-\pi(\omega_1-\omega_2)/a\right]\\\notag
    &\hspace{-1cm}\left.+\exp\left[-\frac{\sigma_A^2}{2}(\Omega_A-\omega_1-\omega_2)^2-\frac{\sigma_B^2}{2}(\Omega_B-\omega_1-\omega_2)^2+i(\Omega_B-\omega_1-\omega_2)\tau_0-\pi(\omega_1+\omega_2)/a\right]\right\}\\\notag
    =&\frac{(2\pi)^3(2\pi a^2)^2}{4^2(4\pi^4a)^2(a^2x_0^2+4)^4}\int_0^\infty d\omega_1\int_0^\infty d\omega_2\frac{1}{\sinh(\pi\omega_1/a)\sinh(\pi\omega_2/a)}\\\notag
    &\times\left[(a^2x_0^2+4)\omega_1 \cos\frac{2\omega_1\text{arcsinh}(ax_0/2)}{a}+\frac{2}{x_0}\sqrt{a^2x_0^2+4} \sin\frac{2\omega_1\text{arcsinh}(ax_0/2)}{a}\right]\\\notag
    &\times\left[(a^2x_0^2+4)\omega_2 \cos\frac{2\omega_2\text{arcsinh}(ax_0/2)}{a}+\frac{2}{x_0}\sqrt{a^2x_0^2+4} \sin\frac{2\omega_2\text{arcsinh}(ax_0/2)}{a}\right]\\\notag
    &\hspace{-1cm}\times\left\{\exp\left[-\frac{\sigma_A^2}{2}(\omega_1+\omega_2+\Omega_A)^2-\frac{\sigma_B^2}{2}(\omega_1+\omega_2+\Omega_B)^2+i(\omega_1+\omega_2+\Omega_B)\tau_0+\pi(\omega_1+\omega_2)/a\right]\right.\\\notag
    &\hspace{-1cm}+\exp\left[-\frac{\sigma_A^2}{2}(\Omega_A+\omega_1-\omega_2)^2-\frac{\sigma_B^2}{2}(\Omega_B+\omega_1-\omega_2)^2+i(\Omega_B+\omega_1-\omega_2)\tau_0+\pi(\omega_1-\omega_2)/a\right]\\\notag
    &\hspace{-1cm}+\exp\left[-\frac{\sigma_A^2}{2}(\Omega_A-\omega_1+\omega_2)^2-\frac{\sigma_B^2}{2}(\Omega_B-\omega_1+\omega_2)^2+i(\Omega_B-\omega_1+\omega_2)\tau_0-\pi(\omega_1-\omega_2)/a\right]\\
    &\hspace{-1cm}\left.+\exp\left[-\frac{\sigma_A^2}{2}(\Omega_A-\omega_1-\omega_2)^2-\frac{\sigma_B^2}{2}(\Omega_B-\omega_1-\omega_2)^2+i(\Omega_B-\omega_1-\omega_2)\tau_0-\pi(\omega_1+\omega_2)/a\right]\right\}.
\end{align}

\section{Calculation of $M$}
\label{App: Calculation of $M$}
In this section we display the results of the calculation.

\begin{align}
\notag
     D_\Psi^i=&\int_0^\infty d\omega_1\int_0^\infty d\omega_2\iint d^2\vec{k}_{1\bot}
    \iint d^2\vec{k}_{2\bot} \frac{1}{1-e^{-2\pi\omega_1/a}}\frac{1}{1-e^{-2\pi\omega_2/a}}\\\notag
    &\times\Bigg\{\theta(\tau_1-\tau_2)e^{-2a\xi_1}\Bigg[\left(\partial_{i_1}v^R_{1\omega_1\vec{k}_{1\bot}}\right)\left(\partial_{i_1}v^R_{1\omega_2\vec{k}_{2\bot}}\right)v^{R*}_{2\omega_1\vec{k}_{1\bot}}v^{R*}_{2\omega_2\vec{k}_{2\bot}} \\\notag
    &+\left(\partial_{i_1}v^R_{1\omega_1\vec{k}_{1\bot}}\right)\left(\partial_{i_1}v^{R*}_{1\omega_2\vec{k}_{2\bot}}\right)v^{R*}_{2\omega_1\vec{k}_{1\bot}}v^R_{2\omega_2\vec{k}_{2\bot}}e^{-2\pi\omega_2/a} \\\notag
    &+\left(\partial_{i_1}v^{R*}_{1\omega_1\vec{k}_{1\bot}}\right)\left(\partial_{i_1}v^R_{1\omega_2\vec{k}_{2\bot}}\right)v^R_{2\omega_1\vec{k}_{1\bot}}v^{R*}_{2\omega_2\vec{k}_{2\bot} }e^{-2\pi\omega_1/a} \\\notag
    &+\left(\partial_{i_1}v^{R*}_{1\omega_1\vec{k}_{1\bot}}\right)\left(\partial_{i_1}v^{R*}_{1\omega_2\vec{k}_{2\bot}}\right)v^R_{2\omega_1\vec{k}_{1\bot}}v^R_{2\omega_2\vec{k}_{2\bot}}e^{-2\pi(\omega_1+\omega_2)/a}\Bigg]\\\notag
    &+\theta(\tau_2-\tau_1)e^{-2a\xi_2}\Bigg[\left(\partial_{i_2}v^R_{2\omega_1\vec{k}_{1\bot}}\right)\left(\partial_{i_2}v^R_{2\omega_2\vec{k}_{2\bot}}\right)v^{R*}_{1\omega_1\vec{k}_{1\bot}}v^{R*}_{1\omega_2\vec{k}_{2\bot}} \\\notag
    &+\left(\partial_{i_2}v^R_{2\omega_1\vec{k}_{1\bot}}\right)\left(\partial_{i_2}v^{R*}_{2\omega_2\vec{k}_{2\bot}}\right)v^{R*}_{1\omega_1\vec{k}_{1\bot}}v^R_{1\omega_2\vec{k}_{2\bot}}e^{-2\pi\omega_2/a} \\\notag
    &+\left(\partial_{i_2}v^{R*}_{2\omega_1\vec{k}_{1\bot}}\right)\left(\partial_{i_2}v^R_{2\omega_2\vec{k}_{2\bot}}\right)v^R_{1\omega_1\vec{k}_{1\bot}}v^{R*}_{1\omega_2\vec{k}_{2\bot} }e^{-2\pi\omega_1/a} \\
    &+\left(\partial_{i_2}v^{R*}_{2\omega_1\vec{k}_{1\bot}}\right)\left(\partial_{i_2}v^{R*}_{2\omega_2\vec{k}_{2\bot}}\right)v^R_{1\omega_1\vec{k}_{1\bot}}v^R_{1\omega_2\vec{k}_{2\bot}}e^{-2\pi(\omega_1+\omega_2)/a}\Bigg]\Bigg\},
\end{align}

\begin{align}
\notag
     D_\Psi^{\bot}=&\int_0^\infty d\omega_1\int_0^\infty d\omega_2\iint d^2\vec{k}_{1\bot}
    \iint d^2\vec{k}_{2\bot} \frac{1}{1-e^{-2\pi\omega_1/a}}\frac{1}{1-e^{-2\pi\omega_2/a}}\\\notag
    &\times\Bigg\{\theta(\tau_1-\tau_2)\Bigg[\left(\nabla_{x_{1\bot}}v^R_{1\omega_1\vec{k}_{1\bot}}\right)\left(\nabla_{x_{1\bot}}v^R_{1\omega_2\vec{k}_{2\bot}}\right)v^{R*}_{2\omega_1\vec{k}_{1\bot}}v^{R*}_{2\omega_2\vec{k}_{2\bot}} \\\notag
    &+\left(\nabla_{x_{1\bot}}v^R_{1\omega_1\vec{k}_{1\bot}}\right)\left(\nabla_{x_{1\bot}}v^{R*}_{1\omega_2\vec{k}_{2\bot}}\right)v^{R*}_{2\omega_1\vec{k}_{1\bot}}v^R_{2\omega_2\vec{k}_{2\bot}}e^{-2\pi\omega_2/a} \\\notag
    &+\left(\nabla_{x_{1\bot}}v^{R*}_{1\omega_1\vec{k}_{1\bot}}\right)\left(\nabla_{x_{1\bot}}v^R_{1\omega_2\vec{k}_{2\bot}}\right)v^R_{2\omega_1\vec{k}_{1\bot}}v^{R*}_{2\omega_2\vec{k}_{2\bot} }e^{-2\pi\omega_1/a} \\\notag
    &+\left(\nabla_{x_{1\bot}}v^{R*}_{1\omega_1\vec{k}_{1\bot}}\right)\left(\nabla_{x_{1\bot}}v^{R*}_{1\omega_2\vec{k}_{2\bot}}\right)v^R_{2\omega_1\vec{k}_{1\bot}}v^R_{2\omega_2\vec{k}_{2\bot}}e^{-2\pi(\omega_1+\omega_2)/a}\Bigg]\\\notag
    &+\theta(\tau_2-\tau_1)\Bigg[\left(\nabla_{x_{2\bot}}v^R_{2\omega_1\vec{k}_{1\bot}}\right)\left(\nabla_{x_{2\bot}}v^R_{2\omega_2\vec{k}_{2\bot}}\right)v^{R*}_{1\omega_1\vec{k}_{1\bot}}v^{R*}_{1\omega_2\vec{k}_{2\bot}} \\\notag
    &+\left(\nabla_{x_{2\bot}}v^R_{2\omega_1\vec{k}_{1\bot}}\right)\left(\nabla_{x_{2\bot}}v^{R*}_{2\omega_2\vec{k}_{2\bot}}\right)v^{R*}_{1\omega_1\vec{k}_{1\bot}}v^R_{1\omega_2\vec{k}_{2\bot}}e^{-2\pi\omega_2/a} \\\notag
    &+\left(\nabla_{x_{2\bot}}v^{R*}_{2\omega_1\vec{k}_{1\bot}}\right)\left(\nabla_{x_{2\bot}}v^R_{2\omega_2\vec{k}_{2\bot}}\right)v^R_{1\omega_1\vec{k}_{1\bot}}v^{R*}_{1\omega_2\vec{k}_{2\bot} }e^{-2\pi\omega_1/a} \\
    &+\left(\nabla_{x_{2\bot}}v^{R*}_{2\omega_1\vec{k}_{1\bot}}\right)\left(\nabla_{x_{2\bot}}v^{R*}_{2\omega_2\vec{k}_{2\bot}}\right)v^R_{1\omega_1\vec{k}_{1\bot}}v^R_{1\omega_2\vec{k}_{2\bot}}e^{-2\pi(\omega_1+\omega_2)/a}\Bigg]\Bigg\},
\end{align}

\begin{align}\notag
W^{\tau1}_{\omega_1\vec{k}_{1\bot},\omega_2\vec{k}_{2\bot}}=&-\omega_1\omega_2\frac{\sinh(\pi\omega_1/a)\sinh(\pi\omega_2/a)}{(4\pi^4a)^2/2\pi}K_{i\omega_1/a}\left(\frac{\kappa_2}{a}\right)^2K_{i\omega_2/a}\left(\frac{\kappa_2}{a}\right)^2 e^{-\sigma^2\Omega^2}\\
&\times e^{i(k_{1x}-k_{2x})x_{0}}e^{-\sigma^2(\omega_1-\omega_2)^2}\Big\{\text{erfc}\left[i\sigma(\omega_1-\omega_2)\right]-2\Big\},\\\notag
W^{\tau2}_{\omega_1\vec{k}_{1\bot},\omega_2\vec{k}_{2\bot}}=&\omega_1\omega_2\frac{\sinh(\pi\omega_1/a)\sinh(\pi\omega_2/a)}{(4\pi^4a)^2/2\pi}K_{i\omega_1/a}\left(\frac{\kappa_2}{a}\right)^2K_{i\omega_2/a}\left(\frac{\kappa_2}{a}\right)^2 e^{-\sigma^2\Omega^2}\\
&\times e^{i(k_{1x}-k_{2x})x_{0}}e^{-\sigma^2(\omega_1-\omega_2)^2}\text{erfc}\left[i\sigma(\omega_1-\omega_2)\right],\\\notag
W^{\tau12}_{\omega_1\vec{k}_{1\bot},\omega_2\vec{k}_{2\bot}}=&\omega_1\omega_2\frac{\sinh(\pi\omega_1/a)\sinh(\pi\omega_2/a)}{(4\pi^4a)^2/2\pi}K_{i\omega_1/a}\left(\frac{\kappa_2}{a}\right)^2K_{i\omega_2/a}\left(\frac{\kappa_2}{a}\right)^2 e^{-\sigma^2\Omega^2}\\
&\times e^{i(k_{1x}+k_{2x})x_{0}}e^{-\sigma^2(\omega_1+\omega_2)^2}\Big\{\text{erfc}\left[i\sigma(\omega_1+\omega_2)\right]-2\Big\}.
\end{align}

\begin{align}\notag
  W^{x_\bot0}_{\omega_1\vec{k}_{1\bot},\omega_2\vec{k}_{2\bot}}=&-\vec{k}_{1\bot}\cdot\vec{k}_{2\bot}\frac{\sinh(\pi\omega_1/a)\sinh(\pi\omega_2/a)}{(4\pi^4a)^2/2\pi}K_{i\omega_1/a}\left(\frac{\kappa_2}{a}\right)^2K_{i\omega_2/a}\left(\frac{\kappa_2}{a}\right)^2 e^{-\sigma^2\Omega^2}\\
&\times e^{i(k_{1x}-k_{2x})x_{0}}e^{-\sigma^2(\omega_1-\omega_2)^2}\Big\{\text{erfc}\left[i\sigma(\omega_1-\omega_2)\right]-2\Big\},\\  \notag    W^{x_\bot1}_{\omega_1\vec{k}_{1\bot},\omega_2\vec{k}_{2\bot}}=&-\vec{k}_{1\bot}\cdot\vec{k}_{2\bot}\frac{\sinh(\pi\omega_1/a)\sinh(\pi\omega_2/a)}{(4\pi^4a)^2/2\pi}K_{i\omega_1/a}\left(\frac{\kappa_2}{a}\right)^2K_{i\omega_2/a}\left(\frac{\kappa_2}{a}\right)^2 e^{-\sigma^2\Omega^2}\\
&\times e^{i(k_{1x}-k_{2x})x_{0}}e^{-\sigma^2(\omega_1-\omega_2)^2}\Big\{\text{erfc}\left[i\sigma(\omega_1-\omega_2)\right]-2\Big\},\\\notag
W^{x_\bot2}_{\omega_1\vec{k}_{1\bot},\omega_2\vec{k}_{2\bot}}=&\vec{k}_{1\bot}\cdot\vec{k}_{2\bot}\frac{\sinh(\pi\omega_1/a)\sinh(\pi\omega_2/a)}{(4\pi^4a)^2/2\pi}K_{i\omega_1/a}\left(\frac{\kappa_2}{a}\right)^2K_{i\omega_2/a}\left(\frac{\kappa_2}{a}\right)^2 e^{-\sigma^2\Omega^2}\\
&\times e^{i(k_{1x}-k_{2x})x_{0}}e^{-\sigma^2(\omega_1-\omega_2)^2}\text{erfc}\left[i\sigma(\omega_1-\omega_2)\right],\\\notag
W^{x_\bot12}_{\omega_1\vec{k}_{1\bot},\omega_2\vec{k}_{2\bot}}=&\vec{k}_{1\bot}\cdot\vec{k}_{2\bot}\frac{\sinh(\pi\omega_1/a)\sinh(\pi\omega_2/a)}{(4\pi^4a)^2/2\pi}K_{i\omega_1/a}\left(\frac{\kappa_2}{a}\right)^2K_{i\omega_2/a}\left(\frac{\kappa_2}{a}\right)^2 e^{-\sigma^2\Omega^2}\\
&\times e^{i(k_{1x}+k_{2x})x_{0}}e^{-\sigma^2(\omega_1+\omega_2)^2}\Big\{\text{erfc}\left[i\sigma(\omega_1+\omega_2)\right]-2\Big\}.
\end{align}

\begin{align}\notag
W^{\xi1}_{\omega_1\vec{k}_{1\bot},\omega_2\vec{k}_{2\bot}}=&-\frac{\kappa_1\kappa_2 \sinh(\pi\omega_1/a)\sinh(\pi\omega_2/a)}{4(4\pi^4a)^2/2\pi}\Big[K_{i\omega_1/a-1}\left(\frac{\kappa_1}{a}\right) +K_{i\omega_1/a+1}\left(\frac{\kappa_1}{a}\right)\Big]\\\notag
&\times \Big[K_{i\omega_2/a-1}\left(\frac{\kappa_2}{a}\right) +K_{i\omega_2/a+1}\left(\frac{\kappa_2}{a}\right)\Big]K_{i\omega_1/a}\left(\frac{\kappa_1}{a}\right) K_{i\omega_2/a}\left(\frac{\kappa_2}{a}\right)\\
&\times e^{-\sigma^2\Omega^2}e^{i(k_{1x}-k_{2x})x_{0}}e^{-\sigma^2(\omega_1-\omega_2)^2}\left\{\text{erfc}\left[i\sigma(\omega_1-\omega_2)\right]-2\right\},\\\notag
W^{\xi2}_{\omega_1\vec{k}_{1\bot},\omega_2\vec{k}_{2\bot}}=&\frac{\kappa_1\kappa_2 \sinh(\pi\omega_1/a)\sinh(\pi\omega_2/a)}{4(4\pi^4a)^2/2\pi}\Big[K_{i\omega_1/a-1}\left(\frac{\kappa_1}{a}\right) +K_{i\omega_1/a+1}\left(\frac{\kappa_1}{a}\right)\Big]\\\notag
&\times \Big[K_{i\omega_2/a-1}\left(\frac{\kappa_2}{a}\right) +K_{i\omega_2/a+1}\left(\frac{\kappa_2}{a}\right)\Big]K_{i\omega_1/a}\left(\frac{\kappa_1}{a}\right) K_{i\omega_2/a}\left(\frac{\kappa_2}{a}\right)\\
&\times e^{-\sigma^2\Omega^2}e^{i(k_{1x}-k_{2x})x_{0}}e^{-\sigma^2(\omega_1-\omega_2)^2}\text{erfc}\left[i\sigma(\omega_1-\omega_2)\right],\\\notag
W^{\xi12}_{\omega_1\vec{k}_{1\bot},\omega_2\vec{k}_{2\bot}}=&-\frac{\kappa_1\kappa_2 \sinh(\pi\omega_1/a)\sinh(\pi\omega_2/a)}{4(4\pi^4a)^2/2\pi}\Big[K_{i\omega_1/a-1}\left(\frac{\kappa_1}{a}\right) +K_{i\omega_1/a+1}\left(\frac{\kappa_1}{a}\right)\Big]\\\notag
&\times \Big[K_{i\omega_2/a-1}\left(\frac{\kappa_2}{a}\right) +K_{i\omega_2/a+1}\left(\frac{\kappa_2}{a}\right)\Big]K_{i\omega_1/a}\left(\frac{\kappa_1}{a}\right) K_{i\omega_2/a}\left(\frac{\kappa_2}{a}\right)\\
&\times e^{-\sigma^2\Omega^2}e^{i(k_{1x}+k_{2x})x_{0}}e^{-\sigma^2(\omega_1+\omega_2)^2}\left\{\text{erfc}\left[i\sigma(\omega_1+\omega_2)\right]-2\right\}.
\end{align}

Now we use polar coordinates to integrate $k_\bot$.

\begin{align}\notag
M^{\Psi\tau}=&\frac{2\pi e^{-\sigma^2\Omega^2}}{(4\pi^4a)^2}\int_0^\infty d\omega_1\int_0^\infty d\omega_2\iint d^2\vec{k}_{1\bot}
    \iint d^2\vec{k}_{2\bot} \frac{\sinh{(\pi \omega_1/a)}\sinh{(\pi \omega_1/a})}{(1-e^{-2\pi\omega_1/a})(1-e^{-2\pi\omega_2/a})}\\\notag
    &\times K_{i\omega_1/a}\left(\frac{\kappa_1}{a}\right)^2K_{i\omega_2/a}\left(\frac{\kappa_2}{a}\right)^2\\\notag
    &\times \Bigg\{-\omega_1\omega_2 e^{i(k_{1x}+k_{2x})x_{0}}e^{-\sigma^2(\omega_1+\omega_2)^2}\text{erfc}\left[i\sigma(\omega_1+\omega_2)\right]\\\notag
    &-\omega_1\omega_2e^{i(k_{1x}-k_{2x})x_{0}}e^{-\sigma^2(\omega_1-\omega_2)^2}\left[\text{erfc}\left[i\sigma(\omega_1-\omega_2)\right]-2\right]e^{-2\pi\omega_1/a} \\\notag
    &+\omega_1\omega_2 e^{i(k_{1x}-k_{2x})x_{0}}e^{-\sigma^2(\omega_1-\omega_2)^2}\text{erfc}\left[i\sigma(\omega_1-\omega_2)\right]e^{-2\pi\omega_2/a}  \\\notag
    &+\omega_1\omega_2e^{i(k_{1x}+k_{2x})x_{0}}e^{-\sigma^2(\omega_1+\omega_2)^2}\left[\text{erfc}\left[i\sigma(\omega_1+\omega_2)\right]-2\right]e^{-2\pi(\omega_1+\omega_2)/a}
    \Bigg\}\\\notag
    =&\frac{2\pi e^{-\sigma^2\Omega^2}}{4(4\pi^4a)^2}\int_0^\infty d\omega_1\int_0^\infty d\omega_2\int_0^\infty dk_{1\bot}\int_0^\infty dk_{2\bot}\int_0^{2\pi} d\theta_1\int_0^{2\pi} d\theta_2 k_{1\bot}k_{2\bot}\\\notag
    &\times \omega_1\omega_2 K_{i\omega_1/a}\left(\frac{\kappa_1}{a}\right)^2K_{i\omega_2/a}\left(\frac{\kappa_2}{a}\right)^2\\\notag
    &\times \Bigg\{-e^{ik_{1\bot}x_0\cos{\theta_1}}e^{ik_{2\bot}x_0\cos{\theta_2}}e^{-\sigma^2(\omega_1+\omega_2)^2}\text{erfc}\left[i\sigma(\omega_1+\omega_2)\right]e^{\pi(\omega_1+\omega_2)/a}\\\notag
    &-e^{ik_{1\bot}x_0\cos{\theta_1}}e^{-ik_{2\bot}x_0\cos{\theta_2}}e^{-\sigma^2(\omega_1-\omega_2)^2}\left[\text{erfc}\left[i\sigma(\omega_1-\omega_2)\right]-2\right] e^{\pi(\omega_2-\omega_1)/a}\\\notag
    &+e^{ik_{1\bot}x_0\cos{\theta_1}}e^{-ik_{2\bot}x_0\cos{\theta_2}}e^{-\sigma^2(\omega_1-\omega_2)^2}\text{erfc}\left[i\sigma(\omega_1-\omega_2)\right] e^{\pi(\omega_1-\omega_2)/a}\\\notag
     &+e^{ik_{1\bot}x_0\cos{\theta_1}}e^{ik_{2\bot}x_0\cos{\theta_2}}e^{-\sigma^2(\omega_1+\omega_2)^2}\left[\text{erfc}\left[i\sigma(\omega_1+\omega_2)\right]-2\right] e^{-\pi(\omega_1+\omega_2)/a}
    \Bigg\}\\\notag
    =&\frac{(2\pi)^3 e^{-\sigma^2\Omega^2}}{4(4\pi^4a)^2}\int_0^\infty d\omega_1\int_0^\infty d\omega_2\int_0^\infty dk_{1\bot}\int_0^\infty dk_{2\bot}(k_{1\bot}k_{2\bot})(\omega_1\omega_2)\\\notag
    &\times K_{i\omega_1/a}\left(\frac{\kappa_1}{a}\right)^2K_{i\omega_2/a}\left(\frac{\kappa_2}{a}\right)^2 J_0(k_{1\bot}x_0)J_0(k_{2\bot}x_0)\\\notag
    &\times \Bigg\{-e^{-\sigma^2(\omega_1+\omega_2)^2}\text{erfc}\left[i\sigma(\omega_1+\omega_2)\right]e^{\pi(\omega_1+\omega_2)/a}\\\notag
    &-e^{-\sigma^2(\omega_1-\omega_2)^2}\left[\text{erfc}\left[i\sigma(\omega_1-\omega_2)\right]-2\right] e^{\pi(\omega_2-\omega_1)/a}\\\notag
    &+e^{-\sigma^2(\omega_1-\omega_2)^2}\text{erfc}\left[i\sigma(\omega_1-\omega_2)\right] e^{\pi(\omega_1-\omega_2)/a}\\\notag
     &+e^{-\sigma^2(\omega_1+\omega_2)^2}\left[\text{erfc}\left[i\sigma(\omega_1+\omega_2)\right]-2\right] e^{-\pi(\omega_1+\omega_2)/a}
    \Bigg\}\\\notag
    =&\frac{(2\pi)^3(\pi a)^2 e^{-\sigma^2\Omega^2}}{4(4\pi^4a)^2 x_0^2 (a^2x_0^2+4)}\int_0^\infty d\omega_1\int_0^\infty d\omega_2 \omega_1\omega_2 \frac{\sin{\frac{2\omega_1\text{arcsinh}(\frac{ax_0}{2})}{a}}}{\sinh{(\pi\omega_1/a)}}\frac{\sin{\frac{2\omega_2\text{arcsinh}(\frac{ax_0}{2})}{a}}}{\sinh{(\pi\omega_2/a)}}\\\notag
    &\times \Bigg\{-e^{-\sigma^2(\omega_1+\omega_2)^2}\text{erfc}\left[i\sigma(\omega_1+\omega_2)\right]e^{\pi(\omega_1+\omega_2)/a}\\\notag
    &-e^{-\sigma^2(\omega_1-\omega_2)^2}\left[\text{erfc}\left[i\sigma(\omega_1-\omega_2)\right]-2\right] e^{\pi(\omega_2-\omega_1)/a}\\\notag
    &+e^{-\sigma^2(\omega_1-\omega_2)^2}\text{erfc}\left[i\sigma(\omega_1-\omega_2)\right] e^{\pi(\omega_1-\omega_2)/a}\\
     &+e^{-\sigma^2(\omega_1+\omega_2)^2}\left[\text{erfc}\left[i\sigma(\omega_1+\omega_2)\right]-2\right] e^{-\pi(\omega_1+\omega_2)/a}
    \Bigg\}.
\end{align}

\begin{align}\notag
M^{\Psi\vec{x}_\bot}=&\frac{2\pi e^{-\sigma^2\Omega^2}}{(4\pi^4a)^2}\int_0^\infty d\omega_1\int_0^\infty d\omega_2\iint d^2\vec{k}_{1\bot}
    \iint d^2\vec{k}_{2\bot} \frac{\sinh{(\pi \omega_1/a)}\sinh{(\pi \omega_1/a})}{(1-e^{-2\pi\omega_1/a})(1-e^{-2\pi\omega_2/a})}\\\notag
    &\times K_{i\omega_1/a}\left(\frac{\kappa_1}{a}\right)^2K_{i\omega_2/a}\left(\frac{\kappa_2}{a}\right)^2\\\notag
    &\times \Bigg\{-\vec{k}_{1\bot}\cdot\vec{k}_{2\bot}e^{i(k_{1x}+k_{2x})x_{0}}e^{-\sigma^2(\omega_1+\omega_2)^2}\text{erfc}\left[i\sigma(\omega_1+\omega_2)\right]\\\notag
    &-\vec{k}_{1\bot}\cdot\vec{k}_{2\bot}e^{i(k_{1x}-k_{2x})x_{0}}e^{-\sigma^2(\omega_1-\omega_2)^2}\left[\text{erfc}\left[i\sigma(\omega_1-\omega_2)\right]-2\right]e^{-2\pi\omega_1/a} \\\notag
    &+\vec{k}_{1\bot}\cdot\vec{k}_{2\bot} e^{i(k_{1x}-k_{2x})x_{0}}e^{-\sigma^2(\omega_1-\omega_2)^2}\text{erfc}\left[i\sigma(\omega_1-\omega_2)\right]e^{-2\pi\omega_2/a}  \\\notag
    &+\vec{k}_{1\bot}\cdot\vec{k}_{2\bot}e^{i(k_{1x}+k_{2x})x_{0}}e^{-\sigma^2(\omega_1+\omega_2)^2}\left[\text{erfc}\left[i\sigma(\omega_1+\omega_2)\right]-2\right]e^{-2\pi(\omega_1+\omega_2)/a}
    \Bigg\}\\\notag
    =&\frac{2\pi e^{-\sigma^2\Omega^2}}{4(4\pi^4a)^2}\int_0^\infty d\omega_1\int_0^\infty d\omega_2\int_0^\infty dk_{1\bot}\int_0^\infty dk_{2\bot}\int_0^{2\pi} d\theta_1\int_0^{2\pi} d\theta_2 k_{1\bot}k_{2\bot}\\\notag
    &\times \vec{k}_{1\bot}\cdot\vec{k}_{2\bot}K_{i\omega_1/a}\left(\frac{\kappa_1}{a}\right)^2K_{i\omega_2/a}\left(\frac{\kappa_2}{a}\right)^2\\\notag
    &\times \Bigg\{-e^{ik_{1\bot}x_0\cos{\theta_1}}e^{ik_{2\bot}x_0\cos{\theta_2}}e^{-\sigma^2(\omega_1+\omega_2)^2}\text{erfc}\left[i\sigma(\omega_1+\omega_2)\right]e^{\pi(\omega_1+\omega_2)/a}\\\notag
    &-e^{ik_{1\bot}x_0\cos{\theta_1}}e^{-ik_{2\bot}x_0\cos{\theta_2}}e^{-\sigma^2(\omega_1-\omega_2)^2}\left[\text{erfc}\left[i\sigma(\omega_1-\omega_2)\right]-2\right] e^{\pi(\omega_2-\omega_1)/a}\\\notag
    &+e^{ik_{1\bot}x_0\cos{\theta_1}}e^{-ik_{2\bot}x_0\cos{\theta_2}}e^{-\sigma^2(\omega_1-\omega_2)^2}\text{erfc}\left[i\sigma(\omega_1-\omega_2)\right] e^{\pi(\omega_1-\omega_2)/a}\\\notag
     &+e^{ik_{1\bot}x_0\cos{\theta_1}}e^{ik_{2\bot}x_0\cos{\theta_2}}e^{-\sigma^2(\omega_1+\omega_2)^2}\left[\text{erfc}\left[i\sigma(\omega_1+\omega_2)\right]-2\right] e^{-\pi(\omega_1+\omega_2)/a}
    \Bigg\}\\\notag
    =&\frac{(2\pi)^3 e^{-\sigma^2\Omega^2}}{4(4\pi^4a)^2}\int_0^\infty d\omega_1\int_0^\infty d\omega_2\int_0^\infty dk_{1\bot}\int_0^\infty dk_{2\bot}(k_{1\bot}k_{2\bot})^2\\\notag
    &\times K_{i\omega_1/a}\left(\frac{\kappa_1}{a}\right)^2K_{i\omega_2/a}\left(\frac{\kappa_2}{a}\right)^2 J_1(k_{1\bot}x_0)J_2(k_{2\bot}x_0)\\\notag
    &\times \Bigg\{e^{-\sigma^2(\omega_1+\omega_2)^2}\text{erfc}\left[i\sigma(\omega_1+\omega_2)\right]e^{\pi(\omega_1+\omega_2)/a}\\\notag
    &-e^{-\sigma^2(\omega_1-\omega_2)^2}\left[\text{erfc}\left[i\sigma(\omega_1-\omega_2)\right]-2\right] e^{\pi(\omega_2-\omega_1)/a}\\\notag
    &+e^{-\sigma^2(\omega_1-\omega_2)^2}\text{erfc}\left[i\sigma(\omega_1-\omega_2)\right] e^{\pi(\omega_1-\omega_2)/a}\\\notag
     &-e^{-\sigma^2(\omega_1+\omega_2)^2}\left[\text{erfc}\left[i\sigma(\omega_1+\omega_2)\right]-2\right] e^{-\pi(\omega_1+\omega_2)/a}
    \Bigg\}\\\notag
    =&\frac{(2\pi)^3 e^{-\sigma^2\Omega^2}}{4(4\pi^4a)^2}\int_0^\infty d\omega_1\int_0^\infty d\omega_2 F(\omega_1,x_0,a)F(\omega_2,x_0,a)\\\notag
    &\times \Bigg\{e^{-\sigma^2(\omega_1+\omega_2)^2}\text{erfc}\left[i\sigma(\omega_1+\omega_2)\right]e^{\pi(\omega_1+\omega_2)/a}\\\notag
    &-e^{-\sigma^2(\omega_1-\omega_2)^2}\left[\text{erfc}\left[i\sigma(\omega_1-\omega_2)\right]-2\right] e^{\pi(\omega_2-\omega_1)/a}\\\notag
    &+e^{-\sigma^2(\omega_1-\omega_2)^2}\text{erfc}\left[i\sigma(\omega_1-\omega_2)\right] e^{\pi(\omega_1-\omega_2)/a}\\
     &-e^{-\sigma^2(\omega_1+\omega_2)^2}\left[\text{erfc}\left[i\sigma(\omega_1+\omega_2)\right]-2\right] e^{-\pi(\omega_1+\omega_2)/a}
    \Bigg\}.
\end{align}

\begin{align}\notag
M^{\Psi\xi}=&\frac{2\pi e^{-\sigma^2\Omega^2}}{4(4\pi^4a)^2}\int_0^\infty d\omega_1\int_0^\infty d\omega_2\iint d^2\vec{k}_{1\bot}
    \iint d^2\vec{k}_{2\bot}\frac{\sinh{(\pi \omega_1/a)}\sinh{(\pi \omega_1/a})}{(1-e^{-2\pi\omega_1/a})(1-e^{-2\pi\omega_2/a})}\\\notag
    &\times \Big[K_{i\omega_1/a-1}\left(\frac{\kappa_1}{a}\right) +K_{i\omega_1/a+1}\left(\frac{\kappa_1}{a}\right)\Big]\Big[K_{i\omega_2/a-1}\left(\frac{\kappa_2}{a}\right) +K_{i\omega_2/a+1}\left(\frac{\kappa_2}{a}\right)\Big]\\\notag
    &\times K_{i\omega_1/a}\left(\frac{\kappa_1}{a}\right) K_{i\omega_2/a}\left(\frac{\kappa_2}{a}\right)\\\notag
    &\times \Bigg\{+e^{i(k_{1x}+k_{2x})x_{0}}e^{-\sigma^2(\omega_1+\omega_2)^2}\text{erfc}\left[i\sigma(\omega_1+\omega_2)\right]\\\notag
    &-e^{i(k_{1x}-k_{2x})x_{0}}e^{-\sigma^2(\omega_1-\omega_2)^2}\left[\text{erfc}\left[i\sigma(\omega_1-\omega_2)\right]-2\right]e^{-2\pi\omega_1/a} \\\notag
    &+ e^{i(k_{1x}-k_{2x})x_{0}}e^{-\sigma^2(\omega_1-\omega_2)^2}\text{erfc}\left[i\sigma(\omega_1-\omega_2)\right]e^{-2\pi\omega_2/a}  \\\notag
    &-e^{i(k_{1x}+k_{2x})x_{0}}e^{-\sigma^2(\omega_1+\omega_2)^2}\left[\text{erfc}\left[i\sigma(\omega_1+\omega_2)\right]-2\right]e^{-2\pi(\omega_1+\omega_2)/a}
    \Bigg\}\\\notag
    =&\frac{(2\pi)^3 e^{-\sigma^2\Omega^2}}{16(4\pi^4a)^2}\int_0^\infty d\omega_1\int_0^\infty d\omega_2\int_0^\infty dk_{1\bot}\int_0^\infty dk_{2\bot}(k_{1\bot}k_{2\bot})^2 J_0(k_{1\bot}x_0)J_0(k_{2\bot}x_0)\\\notag
    &\times \Big[K_{i\omega_1/a-1}\left(\frac{k_{1\bot}}{a}\right) +K_{i\omega_1/a+1}\left(\frac{k_{1\bot}}{a}\right)\Big]\Big[K_{i\omega_2/a-1}\left(\frac{k_{2\bot}}{a}\right) +K_{i\omega_2/a+1}\left(\frac{k_{2\bot}}{a}\right)\Big]\\\notag
    &\times K_{i\omega_1/a}\left(\frac{k_{1\bot}}{a}\right) K_{i\omega_2/a}\left(\frac{k_{2\bot}}{a}\right)\\\notag
    &\times \Bigg\{+e^{-\sigma^2(\omega_1+\omega_2)^2}\text{erfc}\left[i\sigma(\omega_1+\omega_2)\right]e^{\pi(\omega_1+\omega_2)/a}\\\notag
    &-e^{-\sigma^2(\omega_1-\omega_2)^2}\left[\text{erfc}\left[i\sigma(\omega_1-\omega_2)\right]-2\right] e^{\pi(\omega_2-\omega_1)/a}\\\notag
    &+e^{-\sigma^2(\omega_1-\omega_2)^2}\text{erfc}\left[i\sigma(\omega_1-\omega_2)\right] e^{\pi(\omega_1-\omega_2)/a}\\\notag
     &-e^{-\sigma^2(\omega_1+\omega_2)^2}\left[\text{erfc}\left[i\sigma(\omega_1+\omega_2)\right]-2\right] e^{-\pi(\omega_1+\omega_2)/a}
    \Bigg\}\\\notag
    =&\frac{(2\pi)^3 e^{-\sigma^2\Omega^2}}{16(4\pi^4a)^2}\frac{(2a^2\pi)^2}{(a^2x_0^2+4)^4}\int_0^\infty d\omega_1\int_0^\infty d\omega_2 \frac{1}{\sinh{(\pi \omega_1/a)}}\frac{1}{\sinh{(\pi \omega_2/a)}}\\\notag
    &\times \left[(a^2x_0^2+4)\omega_1\cos{\frac{2\omega_1\text{arcsinh}(\frac{ax_0}{2})}{a}}+\frac{2}{x_0}\sqrt{a^2x_0^2+4}\sin{\frac{2\omega_1\text{arcsinh}(\frac{ax_0}{2})}{a}}\right]\\\notag
    &\times \left[(a^2x_0^2+4)\omega_2\cos{\frac{2\omega_2\text{arcsinh}(\frac{ax_0}{2})}{a}}+\frac{2}{x_0}\sqrt{a^2x_0^2+4}\sin{\frac{2\omega_2\text{arcsinh}(\frac{ax_0}{2})}{a}}\right]\\\notag
    &\times \Bigg\{+e^{-\sigma^2(\omega_1+\omega_2)^2}\text{erfc}\left[i\sigma(\omega_1+\omega_2)\right]e^{\pi(\omega_1+\omega_2)/a}\\\notag
    &-e^{-\sigma^2(\omega_1-\omega_2)^2}\left[\text{erfc}\left[i\sigma(\omega_1-\omega_2)\right]-2\right] e^{\pi(\omega_2-\omega_1)/a}\\\notag
    &+e^{-\sigma^2(\omega_1-\omega_2)^2}\text{erfc}\left[i\sigma(\omega_1-\omega_2)\right] e^{\pi(\omega_1-\omega_2)/a}\\
     &-e^{-\sigma^2(\omega_1+\omega_2)^2}\left[\text{erfc}\left[i\sigma(\omega_1+\omega_2)\right]-2\right] e^{-\pi(\omega_1+\omega_2)/a}
    \Bigg\}.
\end{align}




\bibliographystyle{JHEP}

\end{document}